%% file: Paper_arXiv_modif1.tex
\documentclass[12pt,a4paper,final]{iopart}
\pdfoutput=1
\def\contentsname{\empty}
\usepackage{graphicx}
\usepackage{cite}
\usepackage{bigfoot}
\usepackage{appendix}
\usepackage[breaklinks=true,colorlinks=true,linkcolor=black,urlcolor=black,citecolor=black,pdfencoding=auto]{hyperref}
\usepackage{amsfonts, amsmath, dsfont,amssymb}
\usepackage{empheq}
\usepackage{bigfoot}
\usepackage{appendix}
\usepackage{bookmark}

\newcommand{\doidoi}[2]{\href{http://dx.doi.org/#1}{#2}}

\def\contentsname{\empty}
\def\epp{\: .}

\renewcommand*{\geq}{\geqslant}
\renewcommand*{\leq}{\leqslant}

\usepackage[usenames, dvipsnames]{color}

\newcommand{\blue}{ }

\newcommand{\bea}{\begin{eqnarray}}
\newcommand{\eea}{\end{eqnarray}}

\newcommand{\be}{\begin{equation}}
\newcommand{\ee}{\end{equation}}

\def\epp{\:.}

\def\Ai{\text{Ai}}

\usepackage{amsthm}

\newtheorem{theorem}{Theorem}[section]

\theoremstyle{remark}
\newtheorem{remark}[theorem]{Remark}

\allowdisplaybreaks

\newcommand{\nn}{{\nonumber}}
\def\be{\begin{equation}}
\def\ee{\end{equation}}

\makeatletter
\def\@fnsymbol#1{\ifcase#1\or \dagger\or \ddagger\or \S\or
   \|\or \P\or ^{+}\or ^{\tsty *}\or \sharp
   \or \dagger\dagger \or \mbox{$\clubsuit$} \else\@ctrerr\fi\relax}
\makeatother

\usepackage{hyperref}
\hypersetup{
    colorlinks=true,       
    linkcolor=blue,          
    citecolor=magenta,        
    filecolor=red,      
    urlcolor=cyan           
}

\newcommand{\JC}{{\mathbb C}}

\newcommand{\JN}{{\mathbb N}}

\newcommand{\JR}{{\mathbb R}}

\begin{document}

\title[Delta-Bose gas on a half-line and the KPZ equation]{Delta-Bose gas on a half-line and the KPZ equation: boundary bound states and unbinding transitions}

\author{Jacopo De Nardis$^2$, Alexandre Krajenbrink$^{1,3}$,  Pierre Le Doussal$^1$ and Thimoth\'ee Thiery$^{1,4}$}
\address{$^1$ Laboratoire de Physique de l’\'Ecole Normale Sup\'erieure, 
ENS, Universit\'e PSL, CNRS, Sorbonne Universit\'e,
Universit\'e Paris-Diderot, Sorbonne Paris Cit\'e,
24 rue Lhomond, 75005 Paris, France}
\address{$^2$ Department of Physics and Astronomy, University of Ghent, 
Krijgslaan 281, 9000 Gent, Belgium}
\address{$^3$ SISSA and INFN, via Bonomea 265, 34136 Trieste, Italy}
\address{$^4$ Instituut voor Theoretische Fysica, KU Leuven, Leuven, Belgium}
\ead{Jacopo.DeNardis@UGent.be}
\ead{krajenbrink@lpt.ens.fr}
\ead{ledou@lpt.ens.fr}

\begin{abstract}
We revisit the Lieb-Liniger model for $n$ bosons in one dimension with attractive delta interaction 
in a half-space $\mathbb{R}^+$ with diagonal boundary conditions. This model is integrable for arbitrary value of $b \in \mathbb{R}$,
the interaction parameter with the boundary. 
We show that its spectrum exhibits a sequence of transitions, as $b$ is decreased from
the hard-wall case $b=+\infty$, with successive appearance of boundary bound states (or boundary modes) which
we fully characterize. We apply these results to study the Kardar-Parisi-Zhang equation for the growth of a one-dimensional interface of height $h(x,t)$, on the half-space with boundary condition
$\partial_x h(x,t)|_{x=0}=b$ and droplet initial condition at the wall. 
We obtain explicit expressions, valid at all time $t$ and arbitrary $b$, 
for the integer exponential (one-point) moments 
of the KPZ height field $\overline{e^{n h(0,t)}}$. From these moments we extract the large time
limit of the probability distribution function (PDF) of the scaled KPZ height function.
It exhibits a phase transition, related to the unbinding to the wall of the equivalent
directed polymer problem, with two phases: (i) unbound for $b>-\frac{1}{2}$ where the PDF is given by the GSE Tracy-Widom distribution (ii) bound for $b<-\frac{1}{2}$, where the PDF is a Gaussian. At the
critical point $b=-\frac{1}{2}$, the PDF is given by the GOE Tracy-Widom distribution. 
\end{abstract}




\section*{Introduction and aim of the paper} \label{Sec:introduction}

\begin{figure}
\centering
\includegraphics[width=13cm]{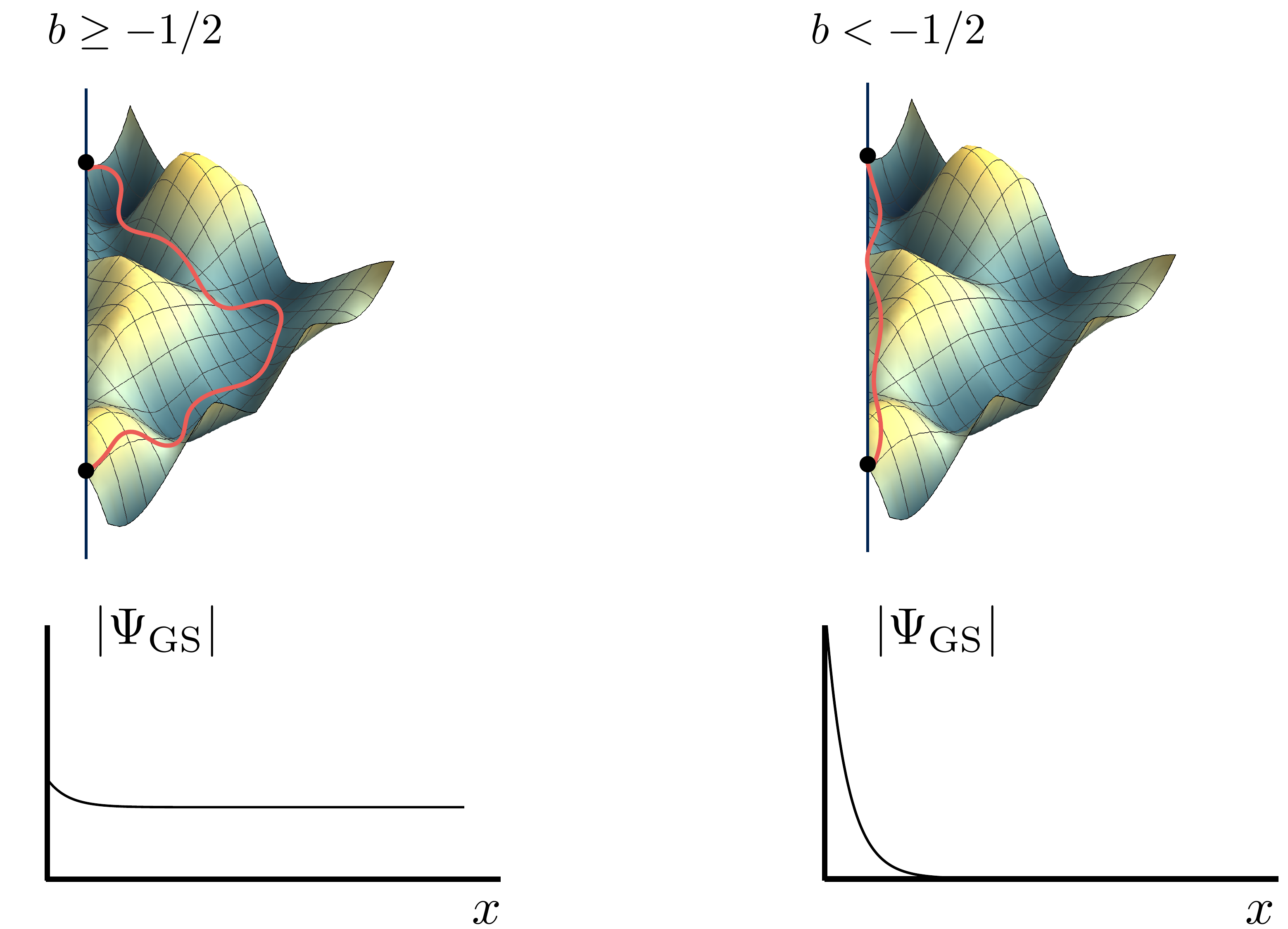}
\caption{\textit{Top}. The directed polymer in random environment with a wall on the left and with fixed endpoints at the wall. The attraction to the wall is parametrized by the coupling $b$. At $b=-1/2$ there is phase transition, for $b<-1/2$ the polymer spend most of the next to the wall, while at $b \geqslant -1/2$ the polymer is unpinned. Depending on the value of $b$ also the statistics of the fluctuations of the free energy (which maps to the KPZ height, see Fig. \ref{fig:kpztimepic}) changes from Gaussian ($b<-1/2$), to GOE Tracy-Widom ($b=-1/2$) and GSE Tracy-Widom ($b>-1/2$). \textit{Bottom}. The same transition is observed for the the ground state wave function of $n$ attractive Lieb-Liniger bosons with open boundary conditions (left wall). The wave function changes from being delocalized ($b \geqslant -1/2$) to being exponentially localized at the boundary ($b <-1/2$) where all particles are in a single bound state bounded to the wall. }
\label{fig:bosons}
\end{figure}

In this paper we study two one-dimensional models in the continuum which play a central role in the current study of out-of-equilibrium many-body systems. The first one is the  Kardar-Parisi-Zhang (KPZ) equation, 
central to the field of stochastic growth,
and the second is the quantum Lieb-Liniger model (LL) for interacting bosons on a line, which represents a paradigmatic model for cold atoms systems. The reason to study both models simultaneously is that the propagator in the LL model, in its attractive regime, maps to the exact solution of the the KPZ equation. In this paper we consider both systems on a half-line (half-space), with open boundary conditions\footnote{in particular we will focus on diagonal open boundary conditions for the XXZ quantum spin models.}.
We will show the existence of boundary multi-particle bound states in the attractive LL model. This is an interesting and new result for the LL model (and non-intuitive
as it also occurs when the boundary is repulsive), which to our knowledge has not been
discussed in details before. In addition, we will obtain solutions for the continuum KPZ equation, or equivalently for the free energy of a continuum directed polymer in random environment, in the
half-space,  which displays a phase transition by changing the values of the coupling with the boundary, see Fig. \ref{fig:bosons}. In the following we first review the recent developments on the KPZ equation and on the LL model.
\subsection*{Overview: KPZ in a half-space.} \label{Sec:intro:kpz}

There has been much recent progress in physics and mathematics in the study of the 1D  (KPZ) universality class, thanks to the discovery of exact solutions and the development of powerful methods to address stochastic integrability and integrable probability. The KPZ  class includes a host of models \cite{HairerSolvingKPZ}: discrete versions of stochastic interface growth such as the PNG growth model \cite{png,spohn2000,ferrari1}, exclusion processes such as the TASEP, the ASEP, the $q$-TASEP and other variants \cite{spohnTASEP,ferrariAiry,Airy1TASEP,ProlhacTASEP,TracyWidom2008,TracyWidom2009,povo,BorodinMacdo,corwinsmallreview}, discrete (i.e. square lattice) \cite{Johansson2000,spohn2000,logsep1,usLogGamma,logsep2,logboro,StrictWeak,StrictWeak2,usIBeta,ThieryStationnary}
or semi-discrete 
\cite{semidiscret1,semidiscret2,OConnellYor} models of directed polymers (DP)
at zero and finite temperature, random walks in time dependent random media
\cite{BarraquandCorwinBeta,PLD-TT-Diffusion}, 
dimer models, random tilings, random permutations 
\cite{LISBaik}, correlation function in quantum condensates \cite{Lamacraft,Altman1,Altman2}, 
and more. 
At the center of this class lies the continuum KPZ equation \cite{KPZ}, see equation \eqref{kpzeq}, which describes the stochastic growth of a continuum interface, and its equivalent formulation in terms of continuum directed polymers (DP)
\cite{directedpoly}, via the Cole-Hopf mapping onto the stochastic heat equation (SHE). Recently exact solutions have also been obtained for the KPZ equation at all times for various
initial conditions
\cite{we,dotsenko,spohnKPZEdge,corwinDP,sineG,we-flat,we-flatlong,PLDCrossoverDropFlat,crissingprob2,dotsenkoGOE,
Quastelflat,SasamotoStationary,BCFV,FerrariSpohnStationary2006,
CorwinLiuWang,BaikFerrariPeche2010,KPZFixedPoint}. 
This was achieved by two different routes. First by studying scaling limits of solvable discrete models, which allowed for rigorous treatments. The second, pioneered by Kardar \cite{kardareplica}, is non-rigorous, but
leads to a more direct solution: it starts from the DP formulation, 
uses the replica method together with a mapping 
to the attractive delta-Bose gas (LL model), which is then studied using the Bethe ansatz (which is nowadays denoted as  replica Bethe ansatz (RBA) method). 
A common aspect of all the models inside the KPZ class is that in the large time limit the KPZ height field (which can be defined for all members of the class) fluctuates on the scale $t^{1/3}$ with $t$ the time since the beginning of the growth. This universality extends beyond scaling: the one point distribution of the field, when appropriately scaled, converges to a limited number of universal distributions that also appear in random matrix theory, the so-called Tracy-Widom distributions for the largest eigenvalue of large Gaussian random matrices \cite{TW1994}. The distribution characterizing the fluctuations of the height at large times depends on some broad features of the initial condition,
i.e. the GUE Tracy Widom distribution for the droplet initial condition, the GOE Tracy Widom  for the flat, and Baik-Rains distribution for the stationary (Brownian). 
These predictions, and a few others, such as multi-point correlations
\cite{ps-npoint,dotsenko2pt,Spohn2ptnew,KPZFixedPoint} and recently
multi-time correlations \cite{deNardisPLDTakeuchi,deNardisPLD2timeLong,PLD2times}, 
have been tested in experiments studying growth between two phases of liquid crystals \cite{exp4,Takeuchi,TakeuchiCrossover,TakeuchiHHLReview}.

\begin{figure}
\centering
\includegraphics[width=10cm]{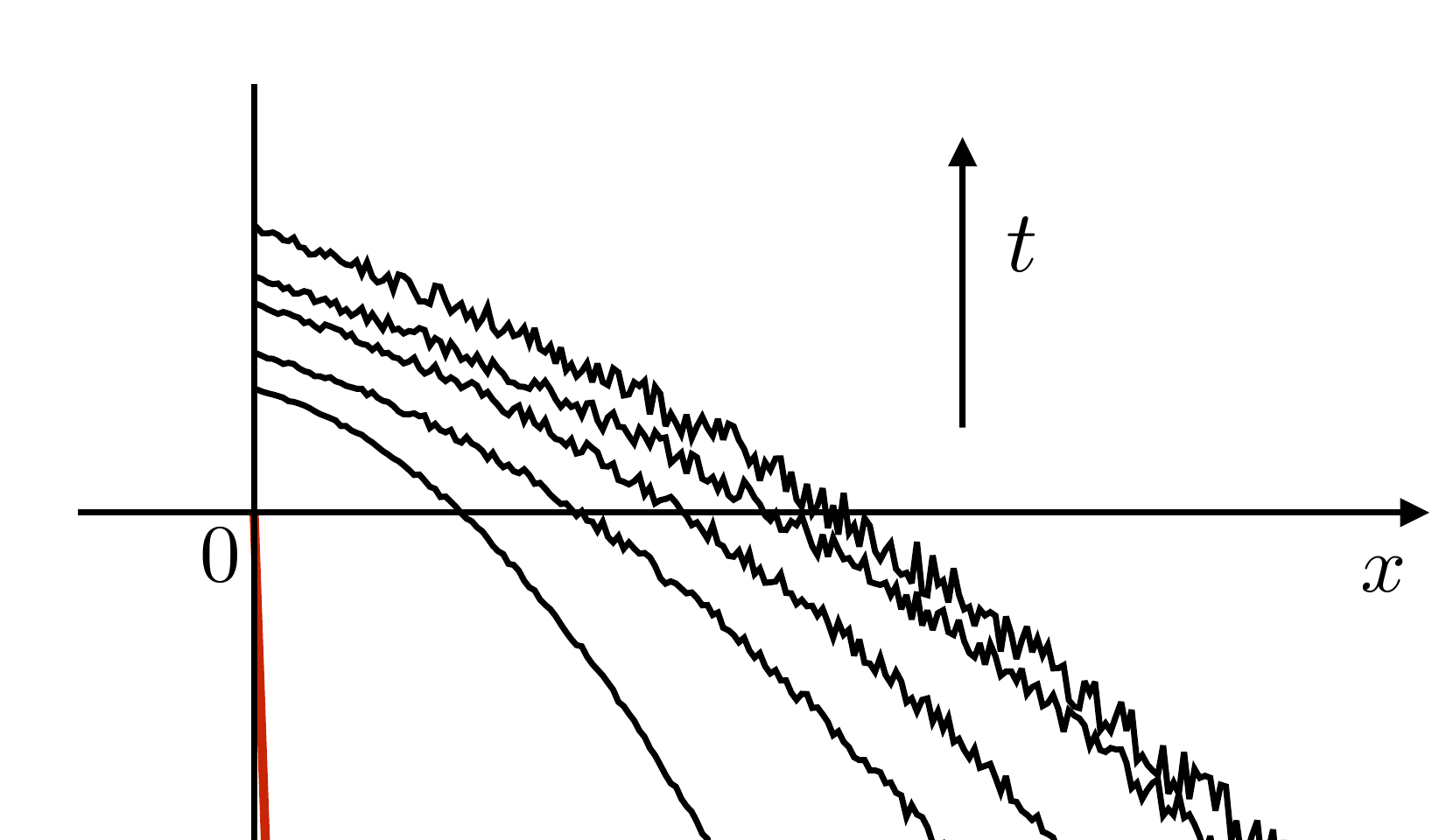}
\caption{Pictorial representation of the time evolution of the KPZ height $h(x,t)$ (black lines) at different times from the droplet initial condition (red line) and with a wall on its left. We here consider the fluctuations of the height $h(x,t)$ next to the wall, namely at $x=0^-$.}
\label{fig:kpztimepic}
\end{figure}
Most of these results have been obtained on the full space. However it is interesting
for applications to study also half-space models, e.g. defined only on the half line $x \in \mathbb{R}^+$. Recently indeed
experiments were able to access the half-space geometries using a bi-regional setting
where the growth rate is different in the two halves of the system
\cite{TakeuchiItoPng}.
Moreover solvability properties (e.g. integrability)are sometimes preserved by going to the half-space, with the proper boundary conditions.
Progress started with discrete models, notably in mathematics. 
In a pioneering paper, Baik and Rains \cite{BaikSymPermutations}
studied the longest increasing sub-sequences (LIS) of {\it symmetrized} random permutations.
The problem maps to a discrete zero temperature model of a directed polymer: one throws
uniformly at random
$2n$ points in a unit square, symmetrically around the diagonal, 
and $m=\alpha \sqrt{2n}$ along the diagonal. Then one looks for the up-right path from the
lowest-leftmost corner to the upper-rightmost one, collecting the maximum number of points,
defined as its "total energy". A second model of a directed polymer was also considered,
with i.i.d. random energies on each site of a square lattice, symmetric around the diagonal. The
weights have geometrical distribution of parameter $q$ in the bulk, and $\alpha \sqrt{q}$ on the
diagonal. These models are the half-space generalizations of two models in the full space,
the LIS of random permutations \cite{LISBaik} and the Johansson DP model \cite{Johansson2000}, 
and allow for an
interaction parameter $\alpha$ with the diagonal (the boundary). Such DP models
can equivalently be seen as stochastic growth models \cite{spohn2000} where the time $t$ corresponds to the
length of the polymer (the number of steps in the second model,
and $\sqrt{n}$ in the first) and the KPZ height to the total energy (a correspondence which
becomes the Cole-Hopf mapping in the continuum, see below). For both models, with the point-to-point
DP geometry, corresponding to the droplet initial condition in the stochastic growth context
(see also \cite{spohn2000}), they found, in the large $t$ limit, a transition when $\alpha$ reaches
the critical value $\alpha_c=1$. For $\alpha<\alpha_c$ the PDF of the
fluctuations of the DP total energy (analog to the height in 
the growth context) is given by Tracy-Widom GSE \cite{TWGSEGOE} on the characteristic KPZ scale $t^{1/3}$. For $\alpha>\alpha_c$
the PDF is Gaussian on the scale $t^{1/2}$, as the DP paths are bound to the diagonal line. 
At the critical point, $\alpha=\alpha_c$ the PDF is given by the GOE Tracy-Widom on the $t^{1/3}$ scale
(note that the GOE also describes the KPZ fluctuations in the full space with flat initial conditions). 
The model thus exhibits a transition from the Gaussian to the KPZ universality class. 
Although the polymer configurations have not, to our knowledge, been studied {in details}
these earlier works, it is clear that, in the DP framework, this is physically a transition between two strategies for optimizing the energy of a path in a random medium: either the path wanders far away from the wall to explore deep valleys of the bulk random potential (unbound phase with KPZ anomalous fluctuations), or the path just follows the diagonal to profit from the diagonal disorder (bound phase pinned to the wall, with Gaussian normal fluctuations). This interpretation was later confirmed in continuum model studies
(see below). In the PNG model in a half-space, with a source
at the origin, a similar transition was found in the PDF of the height at the origin when
the nucleation rate at the origin is increased above a threshold. Below this threshold,
if the height is measured not at the
origin, but at some point away from it, a universal crossover in the fluctuations from the GSE (at the boundary)
to the GUE (in the bulk) Tracy-Widom distribution 
was {shown} \cite{sasamotohalfspace}. The PNG model with two sources
was also studied in \cite{png}. {For the TASEP in a half-space, equivalent to 
last passage percolation in a half-quadrant \cite{SpohnTASEPGSE}, 
similar results were obtained recently using Pfaffian-Schur processes
\cite{BarraquandSchur}
in particular concerning the crossover and the multi-point PDF's.
The case of the asymmetric exclusion process (ASEP) 
was studied in \cite{TWhalf,halfASEPBarraquand}.
Finally, formula were obtained for the (finite temperature) log-gamma DP
with symmetric weights \cite{OConnelSymmetrized,barraquand2018half12}
and other geometries \cite{ZygourasLogGammaHalf}}.

%
%

In physics, the study of the continuum KPZ equation on the half-space was pioneered by Kardar
\cite{KardarTransition}. He considered a continuum DP of length $t$ in a random space-time white noise 
potential on a half space $x>0$, plus a surface potential (represented by an attractive well next
to an impenetrable wall) which amounts to impose $\partial_x h(x,t)|_{x=0}=b$, with a
tunable boundary parameter $b$ (he considered $b<0$ corresponding to an attractive wall). Here
$h(x,t)$ is the equivalent KPZ field which corresponds, via the Cole-Hopf mapping, to minus the free energy of the DP with one endpoint fixed at $x$.
By only considering the ground state of the equivalent half-space LL model, he predicted, 
by heuristic arguments involving the limit of zero number or replica, 
an unbinding transition (that was termed "depinning by quenched randomness") 
from a phase where the DP is bound to the wall for $b <-1/2$, to an unbound phase for
$b>-1/2$, as the attraction to the wall is decreased. His arguments also predict a discontinuous specific heat at the transition,
and an average distance to the wall which diverges as $b \to - \frac{1}{2}^-$.
It is natural to surmise that this transition is the continuum analog of the Baik-Rains (BR) transition of Ref.\cite{BaikSymPermutations} described above, with the critical boundary parameter $b_c=- \frac{1}{2}$ 
corresponding to $\alpha_c=\alpha=1$ in the latter case. In fact, BR obtained a very detailed picture of
this transition, and proved that a universal critical crossover GSE/GOE/Gaussian 
occurs in the fluctuations, on a scale $\alpha -1 \sim t^{-1/3}$ around the transition. These predictions were also found in excellent agreement with numerics
performed on a half-space DP model with non-positive weights \cite{somoza2015unbinding}, demonstrating
that this problem belongs to the same universality class. It suggested a polymer "binding 
length" $t^*$, diverging as $t^* \sim 1/(1-\alpha)^3$ near the transition.
That work also extended the range of applications of this theory to Anderson localization 
in presence of a boundary \cite{somoza2015unbinding}.

An outstanding problem nowadays is to obtain exact solutions for the continuum KPZ equation on a half-space, and for the equivalent continuum DP model. Until recently this has
been achieved only for three values of the parameter $b$. 
In all three cases the solution is for the PDF of the height at (or near) the origin,
for a droplet IC, and for all times. 
For $b=+\infty$, which models a DP with an infinitely repulsive hard wall,
it was obtained in \cite{GueudrePLD} using
the RBA method, and more recently in a distinct, although equivalent form in \cite{krajenbrink2018large}.
Another solution (also non-rigorous) 
was obtained for $b=0$ in \cite{borodin2016directed}, with related but 
different methods using nested contour integral representations,
initiated in \cite{Yudson} in physics and \cite{HeckmanOpdam}
in mathematics. In both cases
the large time limit is found to be GSE-TW, consistent with the general picture of BR
discussed above.
Finally, taking the limit from a half-space ASEP, a rigorous 
solution for the critical case $b=- \frac{1}{2}$ was obtained \cite{halfASEPBarraquand},
leading to GOE-TW.

This problem for a general value of the boundary coupling $b$ has been quite challenging
{(see e.g. discussion in \cite{barraquand2018half12}).}
Very recently, a solution was obtained \cite{AlexLD} using a duality \cite{parekh2019kpz123,parekh2019,footnote0} 
between the droplet IC for any $b$, and the
Brownian IC with a drift $-(b+1/2)$ in presence of an infinite hard wall. The solution is
valid for $b \geq - 1/2$ and for all times, and takes two equivalent forms, in terms of either a matrix kernel or
a scalar kernel. In the large time limit, the GSE-TW distribution is obtained for any $b>-1/2$, while the
GOE-TW distribution is obtained for $b=-1/2$. 

The present paper is a companion paper of Ref. \cite{AlexLD}. Here we do not make use
of the aforementioned duality and study the droplet IC for any value of $b$ directly from the
replica Bethe ansatz in the half-space.
We obtain exact expressions for the $n$-th integer moments 
of the (one-point) DP partition sum (i.e. the exponential of the KPZ field). 
By contrast with Ref. \cite{AlexLD}, they involve an intricate structure of bound states of the half-space
Lieb-Liniger model, which we fully describe here for the first time.
Our expressions for the moments are exact in the whole domain of $n,b$ and $t$ (the corresponding formula without the bound states, valid in a limited domain, were obtained in \cite{PLD1}). Extracting the Laplace transform of the PDF of the DP partition sum 
(the generating function) from these moments is non-trivial, and is performed 
here. In the domain $b \geq -1/2$ we show that the present results agree with the solution obtained
in Ref. \cite{AlexLD}. Although they
are formally valid at all times, they will be analyzed here only in the limit of large time.
That leads to obtaining the universal unbinding phase transition, and 
respectively the GSE, GOE and Gaussian distribution from the RBA
as $b$ is varied. In particular the present method allows to investigate the bound phase (Gaussian
fluctuations).
All our results are achieved by studying the LL model on the half- line, which we now review.

\subsection*{Overview: The attractive Lieb-Liniger model on the half-line with open boundary conditions} \label{Sec:intro:LL}

The Lieb-Liniger model, also called the $\delta$-Bose gas, is a model of spinless bosons
in one dimension interacting with a delta-function potential, see equation \eqref{LL}. It is integrable and its energy spectrum and
eigenfunctions were obtained long ago using the Bethe ansatz \cite{ll}. 
It is an interesting model for quantum gases which has received a revived interest
from experimental realizations in dilute cold atomic gases \cite{exp}. Its properties have been studied in non-equilibrium and equilibrium both theoretically and experimentally, and are quite different in the attractive case and the repulsive case. In the repulsive case, a proper fixed-density thermodynamic limit exists, and
the bosons form a 1D superfluid with quasi long-range order, a sea with particle-hole excitations, and collective modes \cite{ll2} well described at low energy by the bosonisation (i.e. Luttinger liquid) theory \cite{caz}. In the attractive case, on which we focus here, the ground state is a bound state of all the $n$ bosons, and the excitations are obtained by splitting it into a collection of quasi-independent, smaller bound states, which behave almost as free particles, called string states \cite{m-65}. The dynamical correlations have been studied in
\cite{cc-07,muth2010dynamics}. It is an interesting strongly correlated model \cite{bienias2011statistical} with non-trivial bound states, 
also observed in experiments \cite{everitt2015observation,DarkSolitons}. Recently its non-equilibrium properties have received a large attention, in particular after a quantum quench
\cite{piroli2016quantum,Zill,tschischik2015repulsive}.
Connections to the KPZ equation have been discussed in the calculation of overlaps between eigenstates with different value of the coupling \cite{cl-14}
and of the LL quantum propagator \cite{prolhac2011propagator}.

The LL model remains integrable on the half-line with open boundary conditions (also called
mixed in the present context). 
The simplest case is the hardwall case (vanishing wavefunction at the boundary, i.e. $b=+\infty$ in
the present notations, see below)
which was studied in \cite{GaudinHardWall,Ristivojevic2019,TWhalf,LLBoundariesOelkers} 
(see also section 5.1 of \cite{gaudin2014bethe}).
The spectrum of the attractive model was studied (e.g. numerically in \cite{LLChinese})
and it was found that there are still string states as in the bulk,
with no additional bound state associated to the boundary.
There are further integrable generalizations of the LL model associated to reflection 
groups G of $\mathbb{R}^n$ (the so-called generalized kaleidoscope, see
section 5.2 of \cite{gaudin2014bethe}). A large class can be indexed by 
root systems of simple Lie groups \cite{GutkinSutherland,HeckmanOpdam} and 
contain the half-space model studied in this paper,
which has a tunable interaction parameter $b$ with the wall at $x=0$
\cite{gaudin2014bethe,Castillo,borodin2016directed,VanDiejen,EmsizComplete}. 
Here $b$ is an arbitrary real number and the case $b=+\infty$ corresponds to
the hard wall boundary conditions. Note that similar mixed boundary conditions have also been 
studied in the related context of the quantum non-linear Schr\"odinger equation 
\cite{NLSchrod} and in the XXZ spin 1/2 chain \cite{Sklyanin,mailletterras1,mailletterras2,Gohmannopen,Kapustin,denardisTerras}.
However to our knowledge neither in this context,
nor in the LL context, a systematic study of the spectrum for generic value of $b$
has been performed. This is the aim of this paper, and we show
that for any $b<+\infty$ there are boundary bound states that we fully
classify. We will also study the ground state of the system as 
a function of $b$ and of the number of particles $n$. 
%
\newpage
 \section*{Overview of the paper and main results}
The article is divided in few sections which, to certain extent can be read independently. We here review their content and point to the main results.
\begin{itemize}
\item Section \ref{Sec:models} describes the two models that we consider in this work, the Lieb-Liniger model and the KPZ equation, together with its mapping to the DP model, and their half-line versions.
Section \ref{Sec:Models:KPZandLL} explains the connections between the KPZ-DP model and the Lieb-Liniger model on the half-line. 

\item {Section \ref{Sec:LL} provides a complete description of the spectrum of the attractive Lieb-Liniger model on the half-line with the mixed boundary condition defined by Eq. \eqref{bc1}. This model can be also obtained by the scaling limit of an XXZ chain with open boundary conditions and generic longitudinal fields at the boundaries {(see Appendix \ref{app:XXZ})}. 
The generic Bethe eigenfunction for this model is given in \eqref{EqT:wave}. 
Solution of the Bethe equations
show that for any $b<+\infty$ there are two types of eigenstates (i) "bulk strings" as in the full space
(ii) "boundary strings" and that generic eigenstates are build from combination of both.
We show the structure of the ground state as function of $b$ and particle number $n$, in Section \ref{sec:GSLLb}. We obtain the structure of the excitations in Section \ref{sec:LLStringClass}, with a complete characterization of all the possible bound states to the wall (the boundary strings), as a function of the coupling $b$, summarized in Table \ref{tab:stability1} in Section \ref{sec:LLStringClass}. 
Finally we provide an expression for the norm of the Bethe states (eigenstates of the model), Section \ref{sec:normLLsec}, Eq. \eqref{EqT:NormFinal}.}

\item In Section \ref{sec:KPZfullcomp} we then compute the statistics of the fluctuations of the KPZ height 
at one point near the boundary (equivalently, of the directed polymer free energy) for different values of $b$ and for droplet initial conditions. 
In Section \ref{sec:momentsZ} we obtain an exact formula, Eq. \eqref{EqT:SymmetrizedZn},
valid for all time $t$ and all $b$, for the integer moments of the DP partition sum, 
equivalently for the exponential moments $\overline{Z(t)^n} =\overline{e^{n h(0,t)}}$.
In Section \ref{sec:GSandKPZ} we show how some simple properties 
 of the KPZ height function across the transition can be understood simply from the proprieties 
 of the ground state of the bosonic model, Section \ref{sec:GSandKPZ}. In 
 particular we discuss the linear term in the growth profile, $h(0,t) \sim v_\infty(b) t$, and
the form of the velocity $v_\infty(b)$ as function of the coupling $b$.
In Section \ref{sec:defgeneratingfunc} we define the 
convenient generating function of the properly scaled and centered 
KPZ-DP fluctuations, also equal to
the Laplace transform of the DP partition sum, and which identifies with
the CDF of the scaled KPZ height in the large time limit.
In Section \ref{sec:Pfaffian} we show that this generating function
can be written as a Fredholm Pfaffian for all time, and obtain
an exact expression for its kernel for all $b$ and $t$. 
In Section \ref{matching} we show that our result is consistent with
the one of Ref. \cite{AlexLD} and discuss the Mellin-Barnes summation procedure
in Section \eqref{sec:MB}. From then on we only focus on the large time limit.
With the help of the recent results in \cite{AlexLD}, we show in Section \ref{sec:res1} the emergence of: the GSE Tracy-Widom distribution for $b>-1/2$ and the GOE Tracy-Widom distribution for $b=-1/2$.
In Section \ref{sec:attractivebounfdaryGOE} we show the emergence of 
 the Gaussian distribution for $b<-1/2$.
\end{itemize}


{%
	\hypersetup{linkcolor=black}
	\setcounter{tocdepth}{2}
	\tableofcontents
}

\section{Models} \label{Sec:models}

\subsection{The Lieb-Liniger model on a half-line}  \label{Sec:Models:LL}

Consider $n$ identical quantum particles on the line with an attractive interaction
described by the many body Lieb-Liniger (LL) \cite{ll} 
Hamiltonian $H_n$ given by
\be
H_n = -\sum_{j=1}^n \frac{\partial^2}{\partial {x_j^2}}  - 2 \bar c \sum_{1 \leqslant  i<j \leqslant n} \delta(x_i - x_j).
\label{LL}
\ee
and we often denote by a vector $X={x_1,\cdots,x_n}$ the positions of the particles. We 
only study the case of bosons, i.e. $H_n$ acts only on wavefunctions which are fully symmetric 
in the $x_1,\cdots x_n$. On the half-space the
model is defined on $X \in (R^+)^n$, i.e. all $x_i \geqslant 0$, with the following boundary conditions for the wave functions:
\bea \label{bc1}
\partial_{x_i} \Psi(X,t)|_{x_i=0} = b \Psi(X,t)|_{x_i=0}   \quad , \quad i=1\dots ,n \quad , \quad X:= (x_1,\dots,x_n)
\eea 
i.e. the operator $H_n$ acts on the Hilbert space of such functions. \\

To properly define the problem we will consider an additional boundary condition at $x=L$, and be interested
in the limit $L \to +\infty$. Clearly this boundary on the right can be chosen with a generic value of the coupling $b'$. Here we choose $b' \to +\infty$ as this choice should not influence the result 
for any finite $x$ in the large $L$ limit. 

\subsection{The KPZ equation and the directed polymer in a half-space} \label{Sec:Models:KPZDP}

The other model that we study here is the continuum KPZ equation \cite{KPZ}
which describes the stochastic growth of an interface. It is parameterized
by the height $h(x,t)$ at point $x$ evolving as a function of time $t$ as
\be \label{kpzeq}
\partial_t h(x,t) = \nu \partial_x^2 h(x,t) + \frac{\lambda_0}{2}  (\partial_x h(x,t))^2 + \sqrt{D} ~ \xi(x,t)
\ee
driven by a unit white noise $\overline{\xi(x,t) \xi(x',t')}=\delta(x-x') \delta(t-t')$. From now
on we work in the units $x^* = \frac{(2 \nu)^3}{D \lambda_0^2}$, $t^* = \frac{ 2 (2 \nu)^5}{D^2 \lambda_0^4}$ and $h^* = \frac{2 \nu}{\lambda_0}$, so that the KPZ equation becomes
 \bea \label{kpzeq2}
&& \partial_t h(x,t) =  \partial_x^2 h(x,t) + (\partial_x h(x,t))^2 + \sqrt{2} ~ \eta(x,t)  
\eea
where $\eta$ is a unit white noise $\overline{\eta(x,t) \eta(x',t')}=\delta(x-x') \delta(t-t')$.
Here and below overbars denote averages over the white noise $\eta$.
The KPZ equation on the half-line, see Fig. \ref{fig:kpztimepic}, is defined by further restricting  \eqref{kpzeq} to $x>0$
and imposing the boundary condition for all times $t >0$
\be
\partial_x h(x,t)|_{x=0}=b   \label{Pbc} 
\ee 
where the parameter $b \in \mathbb{R}$.\\

On the full space the KPZ equation is mapped,
via the Cole-Hopf transformation $Z(x,t)=e^{h(x,t)}$, to the (multiplicative) 
stochastic heat equation (SHE) 
\bea \label{dp1} 
\partial_t Z(x,t) = \nabla^2 Z(x,t) + \sqrt{2} ~ \eta(x,t) Z(x,t)
\eea 
with Ito convention. It describes a continuous directed polymer (DP) in a quenched random potential, such
that $h(x,t)= \log Z(x,t)$ is (minus) the free energy of the DP of length $t$ with one fixed endpoint 
at $x$. More precisely, for an arbitrary initial condition $h(y,t=0)$, the solution of the KPZ equation
at time $t$ can be written as:
\be \label{ch}
e^{h(x,t)} = Z(x,t) \equiv \int_{-\infty}^{+\infty} \rmd y Z_\eta(x,t|y,0) Z_0(y) \quad , \quad Z_0(y) = e^{h(y,t=0)}.
\ee
Here $Z_\eta(x,t|y,0)$ is the partition function of a continuum directed polymer in the random potential
$- \sqrt{2} ~ \eta(x,t)$ with fixed endpoints at $(y,0)$ and $(x,t)$
\be \label{zdef} 
Z_\eta(x,t|y,0) = \int_{x(0)=y}^{x(t)=x}  \mathcal{D}x(\tau) e^{-  \int_0^t \rmd \tau [ \frac{1}{4}  (\frac{d x}{d\tau})^2  - \sqrt{2} ~ \eta(x(\tau),\tau) ]} 
\ee
where the path integral can be defined as the expectation 
of $e^{ \sqrt{2}  \int_0^t d\tau \eta(x(\tau),\tau)}$
over all Brownian bridges $x(\tau)$ with the
corresponding endpoints. Here $Z_\eta(x,t|y,0)$ also denotes the solution of the 
SHE \eqref{dp1} with initial condition $Z_\eta(x,t=0|y,0)= \delta(x-y)$. 
Equivalently, $Z(x,t)$ is the solution of (\ref{dp1}) with initial conditions $Z(x,t=0)=e^{h(x,t=0)}=Z_0(x)$.\\

On the half-line $x>0$, the same Cole-Hopf transformation $Z(x,t)=e^{h(x,t)}$ can be performed
on Eqs.~\eqref{kpzeq2} and \eqref{Pbc}, and leads to the SHE equation on the half-line \eqref{dp1}
with a Robin (i.e. mixed) boundary condition for $Z(x,t)$, i.e. for any $t \geqslant 0$ and
any $\eta$ one has
\be
\partial_x Z_\eta(x,t)|_{x=0}=b Z_\eta(x=0,t) \label{Pbc2} 
\ee 
We define $Z_\eta(x,t|y,0)$ as the solution of the half-line SHE,
with initial condition $Z_\eta(x,t=0|y,0)= \delta(x-y)$ and $x,y \geqslant 0$.
Then the equation \eqref{ch} also holds on the half-space
with $\int_{-\infty}^{+\infty} \rmd y \to \int_{0}^{+\infty} \rmd y$. 
Furthermore, since $Z_\eta(x,t|y,0)=Z_{\eta'}(y, t|x,0)$ where $\eta'$ is the time reversed environment
$\eta'(x,\tau)=\eta(x,t-\tau)$,
we also have $\partial_y Z_\eta(x,t|y,0)|_{y=0} = b Z_\eta(x,t|y,0)$ for all $t \geqslant 0$ and $\eta$. \\

The propagator $Z_\eta(x,t|y,0)$ on the half-space can be also represented as a sum over directed paths
\be \label{zdef2} 
Z_\eta(x,t|y,0) = \int_{x(0)=y}^{x(t)=x}  \mathcal{D}x(\tau) e^{-  \int_0^t d\tau [ \frac{1}{4}  (\frac{d x}{d\tau})^2  - \sqrt{2} ~ \eta(x(\tau),\tau) + 2 b \, \delta(x(\tau)) ]}, 
\ee
where now the path integral can be defined as an expectation over all {\it reflected}
Brownian bridges $x(\tau) \in \mathbb{R}^+$ with the corresponding endpoints (see discussion in
section 3.2 in \cite{borodin2016directed}). The extra delta interaction
ensures the proper boundary condition at $x=0$.
The parameter $b$ has the dimension of an inverse length and
measures the interaction 
with the wall at $x=0$, $b>0$ being a repulsive wall and $b<0$ attractive. 
The case $b=+\infty$ is the impenetrable (i.e. absorbing) wall. Dividing \eqref{Pbc2} 
by $b$ before taking the limit one sees that it corresponds to
the Dirichlet b.c. , i.e. the partition sum vanishes at the wall, $Z_\eta(0,t|y,0)=0$
and $Z_\eta(x,t|0,0)=0$ for all $\eta, t>0$.
The results for this case have been presented in a short form in \cite{GueudrePLD}.
The case $b=0$ corresponds to the symmetric case and was studied in 
 \cite{borodin2016directed}. 

The droplet initial condition corresponds to the DP with fixed endpoints. Hence
we can adopt the following notation (for the 
solution of the droplet initial condition started in $y$):
\bea \label{drop0} 
h_\eta(x,t|y,0) = \log Z_\eta(x,t|y,0) .
\eea 
In terms of DP the initial conditions is perfectly well defined, while in terms of KPZ height it
requires a short time regularization near $t=0$. Since the calculation is performed 
on the DP framework, it is not important and we will ignore it here.
We will also often omit the "environment" index $\eta$ in further calculations, except when we 
would like to draw attention on the fact that we are considering a fixed realization of the noise (i.e. a given sample). 

\subsection{Connection between the KPZ equation/DP and LL model on the half-line}  \label{Sec:Models:KPZandLL}

As is well known (see e.g.  \cite{kardareplica,bb-00}) the
calculation of the $n$-th integer moment of a DP partition sum can be expressed 
as a quantum mechanical propagator in imaginary time. Since we are studying the continuum KPZ equation
directly with space-time white noise, we connect it to the LL model (i.e. with delta attraction).
In order to show this we consider replicas of (\ref{zdef}), we average over the 
disorder $\eta$ and we use the Feynman-Kac formula. We find that
\be
{\cal Z}_n \equiv  \overline{Z_\eta(x_1,y_1,t)\dots Z_\eta(x_n,y_n,t)}\,,
\ee 
satisfies 
\be
\partial_t {\cal Z}_n = - H_n {\cal Z}_n\,,
\ee 
with $H_n$ the LL Hamiltonian \eqref{LL} with $\bar c = 1$\footnote{In the units that we use in this paper for
the KPZ/DP problem (defined above), the coupling $\bar c$ is set to unity, and $b$ is the only free parameter. The LL model with an arbitrary $\bar c$ corresponds to a DP problem with a partition sum defined as in \eqref{zdef2} but with an additional $\bar c$ in front of the bulk disordered potential $\eta$. Rescaling space $x \to x/\bar c$ and $t \to t/\bar c^2$ in such a a problem brings it back to our units where $b$ here means $b/\bar c$ there.}.
Hence ${\cal Z}_n$ can be written as a quantum mechanical expectation:
\be \label{expect}
{\cal Z}_n = \langle x_1,\dots,x_n | e^{- t H_n} |y_1,\dots,y_n \rangle\,.
\ee
in standard quantum mechanics notations. 
The standard way to compute such an object is to insert the resolution of the identity given by the eigenstates
of $H_n$ and to sum over them.
With arbitrary endpoints $|x_1,..x_n \rangle$ one would need all
eigenstates of $H_n$, not only the symmetric ones (bosonic states). 
However here we need only the moments of the partition sum with {\it fixed endpoints},
hence we need the expectation \eqref{expect} in the states $|x_1,\dots,x_n \rangle= |x,\dots,x \rangle$ and 
$|y_1,\dots,y_n \rangle= |y,\dots,y \rangle$ which are obviously symmetric. These moments
can be written as
\bea \label{EqT:ZnLL1}
\overline{Z_\eta(x,y,t)^n} = \sum_\mu \Psi_\mu(x,\dots,x) \Psi_\mu(y,\dots,y)^* \frac{1}{||\mu||^2} e^{- t E_\mu},
\eea 
i.e. a sum over the unnormalized eigenfunctions $\Psi_\mu$ (of norm denoted $||\mu ||$) 
of $H_{n}$ with energies $E_\mu$ where only symmetric (i.e. bosonic) eigenstates contribute.
Hence we need only to study the LL model, i.e. describing bosons. 
These manipulations were done in the full space, but the case of the half-space is equivalent.
One only has to make sure the boundary conditions are the same,
i.e. \eqref{bc1} is indeed consistent with \eqref{Pbc2}.

Note that the following integrable model on the full line is also often considered 
(see \cite{gaudin2014bethe}
section 5.2) 
\be
H'_n = -\sum_{j=1}^n \frac{\partial^2}{\partial {x_j^2}}  + 2 b \sum_{i=1}^n \delta(x_i) - 2 \bar c \sum_{1 \leqslant  i<j \leqslant n}  \left(\delta(x_i-x_j) + \delta(x_i+x_j) \right), \label{LL2}
\ee 
which corresponds to the continuum directed polymer in a full space with 
a symmetric noise $\eta(x,t)=\eta(-x,t)$ and a delta potential at the origin.
It can be shown to be equivalent to the model considered here,
which lives only on the half-line (see e.g. section 3.2 in \cite{borodin2016directed}
\footnote{The full space model then has initial condition $Z(x,0)=2 \delta(x)$ instead of 
$Z(x,0)=\delta(x)$ see \cite{borodin2016directed}.}).

We will now study in details the properties of the LL model on the half-line obtained by the Bethe ansatz.

\section{The Lieb-Liniger on a half-line: boundary bound states} \label{Sec:LL}

In this section we focus on the Bethe ansatz solution of the attractive Lieb-Liniger model on a half-line $x \in [0, +\infty[$ with a class of integrable boundary conditions at $x=0$ parametrized by $b \in \mathbb{R}$ (the interaction strength with the wall) as already introduced in Sec.~\ref{Sec:Models:LL}. We refer the reader to 
the Introduction for a review of related works on the same model. Our main new contribution consists in a classification of the Bethe roots and states as a funtion of $b$. We introduce a new class of states corresponding to clusters of particles that are bound to the wall (``boundary strings''). To that aim we first study the model on a finite interval $x \in [0,L]$ with the boundary conditions at $L$ taken for convenience as a hard wall (vanishing eigenfunction at $L$) and study the $L \to \infty$ limit of the solutions of the Bethe equations. We obtain an explicit formula for the norms of the string states valid directly in the $L = +\infty$ limit, by studying the $L \to \infty$ limit of the Gaudin formula for the norm within a generalization of the method used in \cite{cc-07}. Then we discuss the ground state and the excitations, and we discuss the quantum phase transition of the ground state as a function of $n$ and $b$.

\subsection{Finite size model}


Here we first consider the quantum mechanics of $n \in \mathbb{N}$ bosons on a line $X=(x_1 , \cdots , x_n) \in [0,L]^n$ with attractive zero-range interaction. We use units such that $\hbar=1$ and the mass of the particle is $m=1/2$. The Schr\"odinger equation for the wave-function is taken as
\bea 
&& i 
\partial_t \psi_t(X) = H_n \psi_t(X) \quad , \quad \text{for $(x_1 , \dots , x_n) \in ]0,L[^n$}  \nn \\
&& \partial_{x_i} \psi_t(X)|_{x_i = 0} = b  \psi_t(X)|_{x_i = 0} \quad , \quad \forall i \in [1,n] \nn \\
&& \psi_t(X)|_{x_i = L} = 0  \quad , \quad \forall i \in [1,n] \label{LLBCdef2} 
\eea
with $H_n$ the attractive Lieb-Liniger Hamiltonian \eqref{LL} that we recall here,
\be  \label{LLdef2} 
H_n = -\sum_{j=1}^n \frac{\partial^2}{\partial {x_j^2}}  - 2 \bar c \sum_{1 \leqslant  i<j \leqslant n} \delta(x_i - x_j) \, ,
\ee
with $\bar c >0$.

\subsection{Bethe ansatz solution: wave-functions and Bethe equations}
~

It is known, see Refs.~\cite{gaudin2014bethe,VanDiejen,LLBoundariesOelkers}, that the above model is solvable by Bethe ansatz: the following spectral problem
\bea \label{Spectral1}
&& H_n \Psi_\mu(X) = E_\mu \Psi_\mu(X)  \quad , \quad \text{for $(x_1 , \dots , x_n) \in ]0,L[^n$} ;  \label{EqT:Spectral1} \\
&& \partial_{x_i} \Psi_\mu(X)|_{x_i = 0} = b  \Psi_\mu(X)|_{x_i = 0} \quad , \quad \forall i \in [1,n] ;  \label{EqT:Spectral2} \\
&& \Psi_\mu(X)|_{x_i = L} = 0  \quad , \quad \forall i \in [1,n]. \label{EqT:Spectral3}
\eea
admits solution of the Bethe form, parametrized by a set of rapidities $(\lambda_1 , \cdots , \lambda_n) \in \mathbb{C}^n$, symmetric in $(x_1, \cdots ,x_n)$ and given by, in the $0 \leqslant x_1 \leqslant  \dots   \leqslant x_n \leqslant L $ sector,
\bea \label{EqT:wave}
&& \!\!\!\!\!\!\!\!\!\!\!\!\!\!\!\!\!\!\!\!\!\!\!\!\!\!\!\!\!\!\!\!\!\!\!\!\!\! \Psi_\mu(x_1 \leqslant  \cdots   \leqslant x_n) = \frac{1}{(2 i)^{n}} \sum_{P \in S_n} \sum_{\epsilon_1 \epsilon_2,..\epsilon_n=\pm 1} \epsilon_1 \epsilon_2\dots \epsilon_n A[\epsilon_1 \lambda_{P_1},\epsilon_2 \lambda_{P_2}, \dots \epsilon_n \lambda_{P_n}]  \exp\left( i \sum_{\alpha=1}^n \epsilon_\alpha \lambda_{P_\alpha} x_\alpha   \right) \nn  \\
&& \!\!\!\!\!\!\!\!\!\!\!\!\!\!\!\!\!\!\!\!\!\!\!\!\!\!\!\!\!\!\!\!\!\!\!\!\!\! A[\lambda_1,..,\lambda_n]= \prod_{1\leqslant \alpha < \beta \leqslant n } (1+ \frac{i \bar c}{\lambda_\beta - \lambda_\alpha})(1+ \frac{i \bar c}{\lambda_\beta + \lambda_\alpha})
\prod_{\alpha=1}^n (1 + i \frac{\lambda_\alpha}{b}) \, .
\eea
where the sum is over permutations of $S_n$.
Such wavefunctions are continuous over $(\mathbb{R}^{+})^n$. 
They automatically satisfy \eqref{EqT:Spectral1}, or equivalently, the matching condition
arising from the $\delta(x_i-x_j)$ interaction:
\be
 (\partial_{x_{i+1}} -  \partial_{x_{i}} + \bar c) \Psi_\mu(x_1 \leqslant \cdots\leqslant  x_n)|_{x_{i+1}=x_i^+}=0 
\ee 
as well as the boundary condition \eqref{EqT:Spectral2}. The condition \eqref{EqT:Spectral3} is enforced by asking that the rapidities satisfy the following Bethe equations
\footnote{All of these properties hold for more general boundary conditions at the second boundary, i.e. $\partial_{x_i} \Psi_\mu(X)|_{x_i = L} = - b'  \Psi_\mu(X)|_{x_i = L}$, the only change being 
an extra factor $\frac{b' - i \lambda_\alpha}{b'+ i \lambda_\alpha}$ in the r.h.s. of \eqref{EqT:BetheEq1}
which is unity for the case $b'=+\infty$ considered here.}
\bea \label{EqT:BetheEq1}
e^{2 i \lambda_\alpha L} = 
\frac{b - i \lambda_\alpha}{b+ i \lambda_\alpha} \, \prod_{1 \leqslant \beta \neq \alpha \leqslant n} \frac{\lambda_\alpha - \lambda_\beta - i \bar c}{\lambda_\alpha - \lambda_\beta + i \bar c} 
\frac{\lambda_\alpha + \lambda_\beta  - i \bar c}{\lambda_\alpha  + \lambda_\beta +  i \bar c},  \,  
 \quad \forall \alpha \in [1,n] \, .
\eea 
It can be directly checked that such solutions indeed solve the spectral problem \eqref{EqT:Spectral1}-\eqref{EqT:Spectral2}-\eqref{EqT:Spectral3} with energy $E_\mu = \sum_{\alpha=1}^n \lambda_\alpha^2$. 
In the following we will construct a set of solutions to the Bethe equations in the $L= \infty$ limit
which we will conjecture to be a complete set.

\begin{remark}
Note that it is obvious from \eqref{EqT:wave} that changing any $\lambda_j \to -\lambda_j$ (for any given $j$) just changes the wavefunction as $\Psi_\mu \to - \Psi_\mu$. This parity symmetry will be taken into account later on when classifying states.
\end{remark}

\begin{remark}
We also note that in the limit $\bar c=0$ (no interaction) the wavefunction \eqref{EqT:wave} becomes:
\be
 \Psi_\mu(x_1,\dots,x_n) = \underset{n \times n}{\rm Perm}  [\sin (\lambda_i x_j) + \frac{\lambda_i}{b} \cos(\lambda_i x_j) ]
\ee
where ${\rm Perm}[A]=\sum_P \prod_{i=1,n} A_{i,P_i}$ is the permanent of the $n \times n$ matrix $A$, 
as expected for a non-interacting bosonic eigenstate.
\end{remark}

\subsection{Large $L$ limit: classification of strings}\label{sec:LLStringClass}
In this section we obtain the classification of the so-called strings solutions to the Bethe equations, generalizing the usual known case of an infinitely repulsive wall $b = +\infty$ (absorbing boundary condition at $x=0$ for the DP problem) that we first recall here.

\subsubsection{In the $b \to +\infty$ limit: bulk strings.}
In that limit the Bethe equations reduce to 

\bea \label{EqT:usualBethe}
e^{2 i \lambda_\alpha L} = \prod_{1 \leqslant \beta \neq \alpha \leqslant n} \frac{\lambda_\alpha - \lambda_\beta - i \bar c}{\lambda_\alpha - \lambda_\beta + i \bar c} 
\frac{\lambda_\alpha + \lambda_\beta  - i \bar c}{\lambda_\alpha  + \lambda_\beta +  i \bar c},   \quad \forall \alpha \in [1,n] \, .
\eea 

It is in that case well known \cite{cc-07,LLChinese} that in the dilute limit $L \to \infty$ with $n$ fixed a complete set of solutions of the Bethe eigenfunctions can be obtained as follows: the set of rapidities $\lambda_{\alpha}$ can be decomposed in $n_{s}$ packets, also called {\it strings}, of $m_{i} \geqslant 1$ particles (the string {\it multiplicities}) with $i=1, \dots , n_s$ and $n = \sum_{i=1}^{n_s} m_j$. Inside the $j$-{th} string, the quasi-momenta are labeled by $a= 1 , \dots , m_j$ and take the form
 \be \label{EqT:UsualString}
\lambda^{j,a} := k_j + \frac{i \bar c}{2} (m_j +1 - 2 a) + \delta^{j a}, 
 \ee
where $\delta^{ja} \sim e^{-\hat{\eta}^{ja} L} $ (with $\hat{\eta}^{ja} >0$) are corrections to this leading behavior that decay exponentially with $L$. For many practical purposes the different strings that compose a given state can be considered as independent and the string `momenta' $k_j$ are independently quantized as $k_j = \frac{I_j \pi}{L}$, $I_j \in \JN$. State with rapidites as in \eqref{EqT:UsualString} will henceforth be referred to as `bulk strings'. Physically these states correspond to bound states with the wavefunction being supported in the fraction of space with the different particles distant from one another by a distance of order $1/\bar c$ and the center of mass uniformly distributed in the bulk. This can seen by looking at the wavefuntion for a single string of $n$ particles ($n_s=1$): $\Psi_{\mu}(x_1,\cdots,x_n) = n! \exp\left(-\frac{\bar c}{2} \sum_{1\leq \alpha < \beta \leqslant n} |x_\alpha - x_\beta| \right)$. The contribution to the energy of the string state with momenta $k_j$ and multiplicity $m_j$ is
\be \label{EqT:energy}
\sum_{a=1}^{m_j} (\lambda^{ja})^2=  m_j k_j^2- \frac{\bar c^2}{12} m_j(m_j^2-1)  =: E(m_j,k_j) 
\ee

\begin{figure}
\centering
\includegraphics[width=5cm]{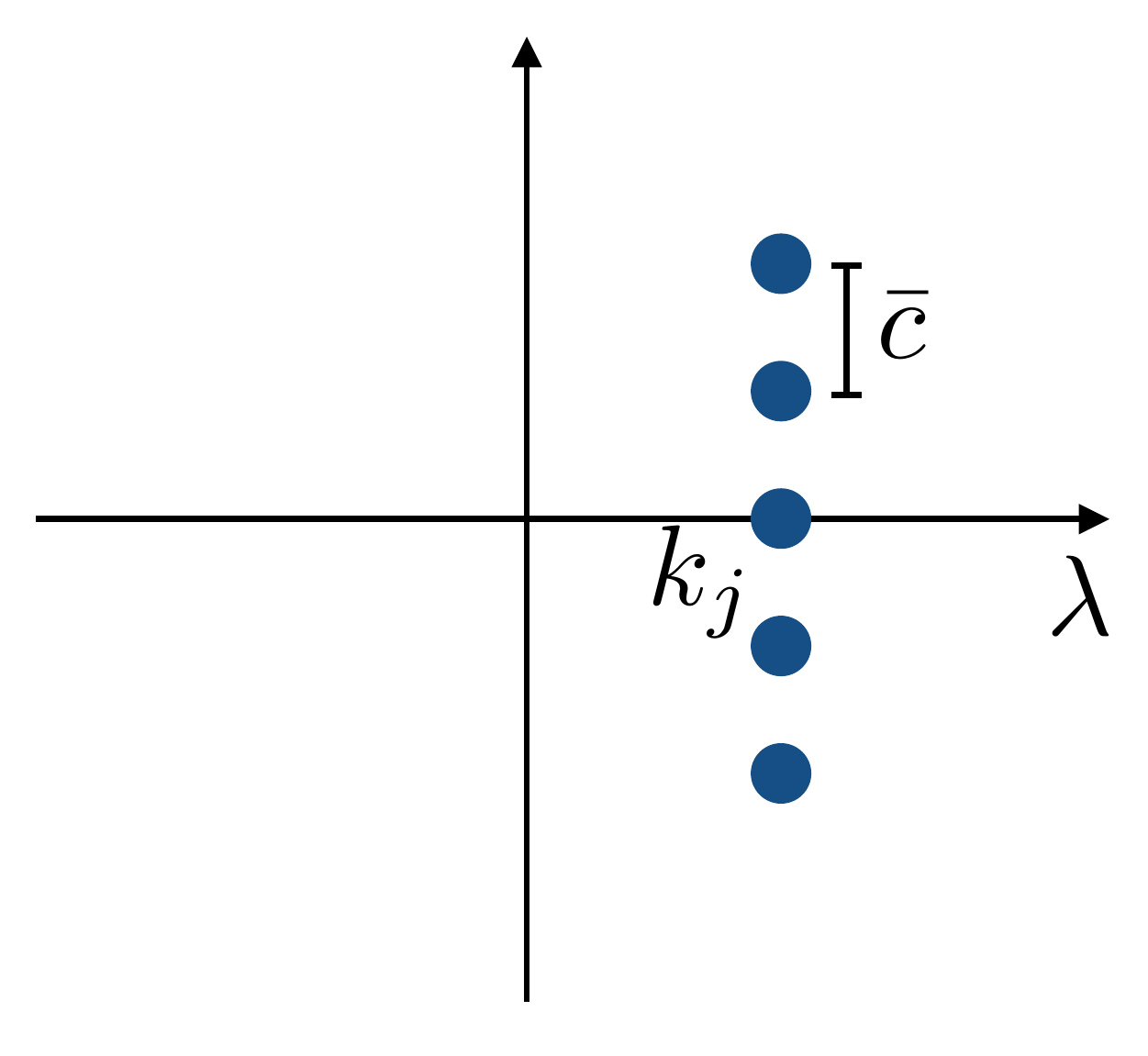}
\caption{Rapidities inside a bulk string given by Eq. \eqref{EqT:UsualString} (here $m_j=5$ and $k_j>0$) are always arranged in a way that is symmetric with respect to the real axis and generally have a non-zero real part (the string momentum $k_j$).}
\label{fig:string1}
\end{figure}

 The consistency of this so-called string solution can be checked by directly inserting \eqref{EqT:UsualString} into \eqref{EqT:usualBethe}. Keeping only the terms that are exponential in $L$ we obtain (i.e. we check the logarithm of \eqref{EqT:usualBethe} in the large $L$ limit), denoting $\delta^{ja} - \delta^{ja+1} \sim e^{-\eta^{aa+1} L}$ for $a=1,\cdots , n$ (i.e. $\eta^{a,a+1} = {\rm min}(\hat{\eta}^{ja}, \hat{\eta}^{ja+1})$) 
\be
 e^{- L \bar c(m_j+1-2a)} \sim e^{-(\eta^{a,a+1}-\eta^{a-1,a} )L},  \quad \text{ for $a \in [1,m_j]$}  \, ,
\ee
where we have introduced $\eta^{0,1}=\eta^{m_j,m_j+1}=0$ for notational simplicity. Solving for the $\eta^{aa+1}$ leads to $\eta^{aa+1} = (m_j-1)\bar c + \bar{c}(a-1)(m_j-1-a)$. These indeed correspond to deviations $\delta$ that decay exponentially to $0$ if the $\eta^{aa+1}$ are positive, which is always the case when $\bar{c}>0$ (attractive interaction). From the analytic point of view notice that these string solutions are possible because of the singularities of the right hand side of the Bethe equations \eqref{EqT:usualBethe} that make possible the existence of a non-zero imaginary part of the rapidites in the large $L$ limit.

\begin{remark}
Note that in constructing these string solutions we only `use' the poles of the first factor in the Bethe equations and the presence of the second factor was inconsequential above. It can be easily seen that `using' the second factor to build up new string solutions is also possible. This will however create string solutions identical to the one introduced above, with possibly a few $\lambda^{ja}$ transformed into $-\lambda^{ja}$. As already remarked above, these solutions thus simply correspond to the same eigenstate and should not be counted twice.
\end{remark}

\subsubsection{Arbitrary $b$: boundary strings.} 

\begin{figure}
\centering
\includegraphics[width=9cm]{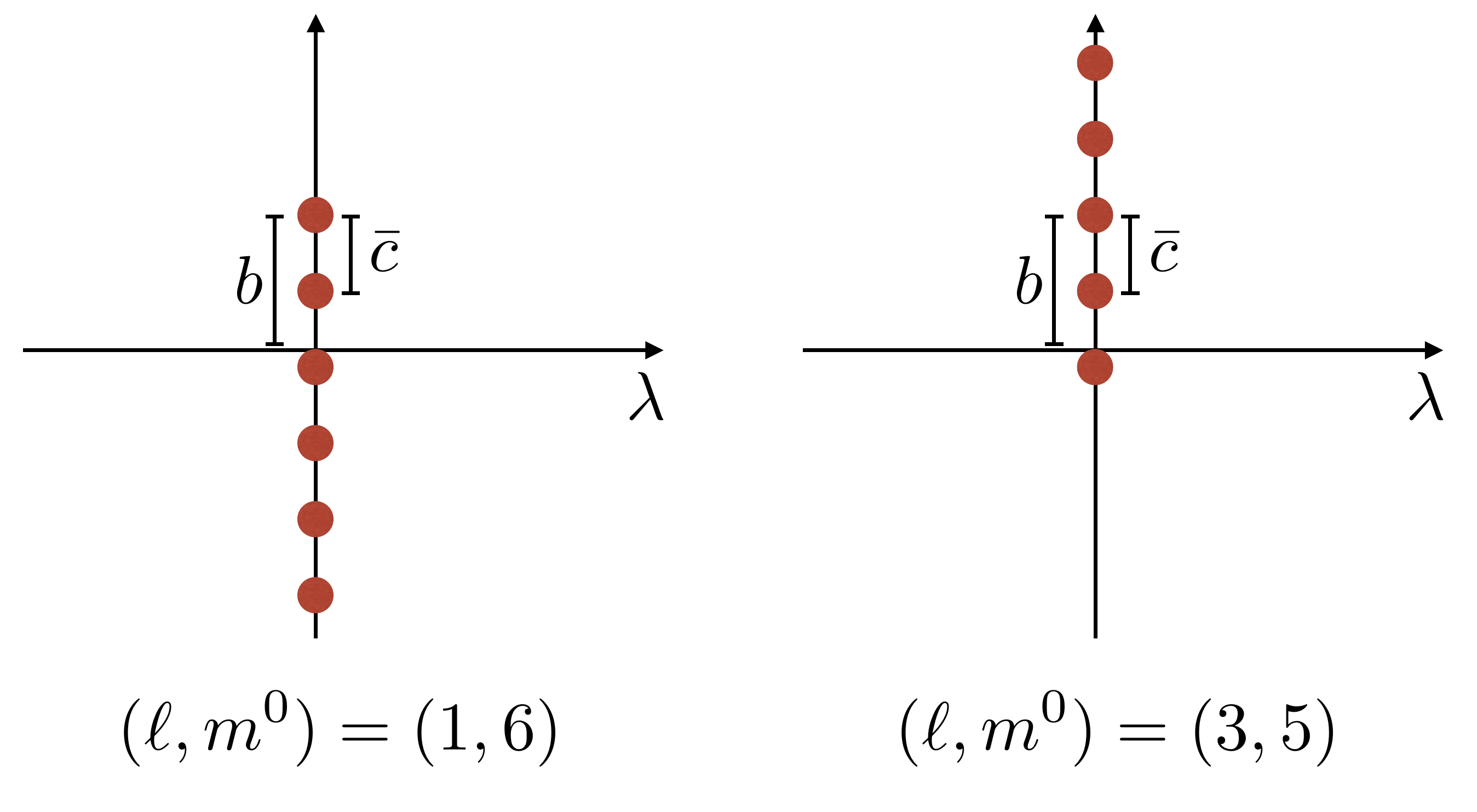}
\caption{Rapidities inside boundary strings described by Eq. \eqref{EqT:BoundaryStringRapidities} are arranged in a way similar as for bulk strings, except that the real part (the real momentum) is always vanishing (states are bound to the wall) and the rapidities are in general not symmetric with repect to the real axis. The integer $\ell$ denotes the position at which the rapidity $\lambda_{a=\ell} = i b $ appears inside the boundary string.}
\label{fig:string2}
\end{figure}

For $b$ finite we rewrite here for clarity the Bethe equations
\bea \label{EqT:BetheEq2}
e^{2 i \lambda_\alpha L} = \prod_{\beta \neq \alpha} \frac{\lambda_\alpha - \lambda_\beta - i \bar c}{\lambda_\alpha - \lambda_\beta + i \bar c} 
\frac{\lambda_\alpha + \lambda_\beta  - i \bar c}{\lambda_\alpha  + \lambda_\beta +  i \bar c}  \,  \frac{b - i \lambda_\alpha}{b+ i \lambda_\alpha}, \quad \forall \alpha \in [1,n] \, .
\eea 
Inspection of the above equations show that the appearance of the last term in \eqref{EqT:BetheEq2} as compared to \eqref{EqT:usualBethe} makes possible the existence of new string states that `use' the singularities of this term by having one rapidity $\lambda_{\alpha}$ equal to $\pm ib$ in the large $L$ limit. Other rapidities can then be added on top by using the singularities originating from the other factors $\lambda_\alpha \pm \lambda_\beta \pm i \bar c$, as for the usual string states. Relying on the symmetry $\lambda_\alpha \to -\lambda_{\alpha}$ of the Bethe states, we can restrict our search for these new states to states with $m^0 \ge 1$ rapidities, which we denote $\lambda^{0a}$ (we generally use the superscript $0$ for
a boundary string)
with
\bea \label{EqT:BoundaryStringRapidities}
\lambda^{0a} := i b + i \bar c (\ell-a) + \delta_0^{\ell,a} \quad , \quad a=1,\dots, m^0.
\eea
with $\delta_0^{\ell,a}$ deviations that decay exponentially to $0$ with $L$ and $\ell \in [1,m^0]$ an integer that parametrize the position of $i b$ in the string (see Fig.~\ref{fig:string2}). A state with rapidities as in \eqref{EqT:BoundaryStringRapidities} will henceforth be referred to as a ``boundary string" $(\ell,m^0)$.

 Writing $\delta_0^{\ell,a} -\delta_0^{\ell,a+1} \sim e^{-2 \eta^{aa+1} L}$ for $a=1,\cdots,m^0-1$ and $\delta_0^{\ell,\ell}  \sim e^{-2\eta^{\ell} L}$ with all $\eta$ symbols positive, we check the consistency of this string solution by inserting \eqref{EqT:BoundaryStringRapidities} into \eqref{EqT:BetheEq2}. We obtain, to logarithmic accuracy,
\bea \label{EqT:SystemEta}
&& e^{-(b+\bar c (\ell-a)) L} \sim e^{-(\eta^{aa+1}-\eta^{a-1a})L}  \times e^{\delta_{a,\ell} \eta^\ell L}\quad , \quad a \in [1,m^0]  \, ,
\eea
with again $\eta^{0,1}=\eta^{m^0,m^0+1}=0$ by convention.
This can be solved and we obtain
\bea \label{EqT:SolutionEta}
&& \!\!\!\!\!\!\!\!\!\!\!\!\!\!\!\!\!\!\!\!\!\!\!\! \eta^{aa+1} = a b +\left( a \ell - \frac{a(a+1)}{2} \right)\bar c \quad  1 \leqslant a \leqslant \ell-1 \nn \\
&& \!\!\!\!\!\!\!\!\!\!\!\!\!\!\!\!\!\!\!\!\!\!\!\! \eta^{aa+1} =-(m^0-a)b + \left( (m^0-\ell)(m^0-a) - \frac{(m^0-1-a)(m^0-a)}{2}   \right) \bar c \quad  \ell \leqslant a \leqslant m^0-1 \nn \\ 
&& \!\!\!\!\!\!\!\!\!\!\!\!\!\!\!\!\!\!\!\!\!\!\!\! \eta^\ell = - m^0 b + \frac{\bar c }{2}\left((m^0-\ell)(m^0-\ell+1)-\ell(\ell-1)\right) \, .
\eea
The system of equations \eqref{EqT:SystemEta} is slightly different if $\ell=1$ or $\ell = m^0$ but the solution \eqref{EqT:SolutionEta} for the $\eta^\ell$ remains identical. The solution \eqref{EqT:SolutionEta} can then be used to conclude on the domain of stability of these solutions: this string solution is consistent with the Bethe equations if all the $\eta$ symbols are positive. Requiring this in general implies a lower bound on $b$ (from the first line of \eqref{EqT:SolutionEta}, that only exists if $\ell \geqslant 2$) and an upper bound on $b$ (from the second and third lines of \eqref{EqT:SolutionEta}). It can be seen that the most restrictive lower bound is obtained for $\ell \geqslant 2$ by requiring that $\eta^{\ell-1\ell}>0$, and the most restrictive upper bound is obtained by requiring that $\eta^{\ell} >0$. From that we obtain that the string state $(\ell,m^0)$ exists if
\bea \label{EqT:Existence}
\text{$\ell=1$ and $b < \frac{\bar c}{2}(m^0-1)$}, \nn \\
\text{$\ell \geqslant 2$ and $- \ell \frac{\bar c }{2} < b < \frac{\bar c}{2}(m^0-2 \ell+1)$} \, .
\eea
We summarize for clarity the possible boundary strings as a function of $b/\bar{c}$ 
in the Table~\ref{tab:stability1}.

\begin{center}
\begin{table}
  \begin{tabular}{| c | c | }
    \hline
   Value of $b$ &  Allowed Boundary Strings   \\ [0.5ex] \hline  \hline 
	$p/2\leqslant \frac{b}{\bar{c}} < (p+1)/2$, $p \in \mathbb{N}$  & $ (\ell\geq 1, m \geqslant 2 \ell +p)$     \\[0.5ex] \hline 
     $3/2\leqslant \frac{b}{\bar{c}} < 2$  & $ (\ell\geq 1, m \geqslant 2 \ell +3)$     \\ [0.5ex] \hline

     $1\leqslant \frac{b}{\bar{c}} < 3/2$  & $ (\ell\geq 1, m \geqslant 2 \ell +2)$     \\ [0.5ex] \hline
     $1/2\leqslant \frac{b}{\bar{c}} < 1$  & $ (\ell\geq 1, m \geqslant 2 \ell +1)$     \\ [0.5ex] \hline

     $0\leqslant \frac{b}{\bar{c}} < 1/2$  & $ (\ell\geq 1, m \geqslant 2 \ell )$     \\ [0.5ex] \hline
     $-1/2\leqslant \frac{b}{\bar{c}} < 0$  & $(\ell \geqslant 1 , m \geqslant 2 \ell -1)$     \\ [0.5ex] \hline
      $-1< \frac{b}{\bar{c}} < -1/2$  & ${\underline{ (\ell= 1, m \geqslant 1 )}}$ and $(\ell \geqslant 2 , m \geqslant 2 \ell -2)$     \\ [0.5ex] \hline
      $\frac{b}{\bar{c}} =-1$  & {$\underline{ (\ell= 1, m \geqslant 1 )}$} and $(\ell \geqslant 3 , m \geqslant 2 \ell -2)$     \\ [0.5ex] \hline
      $-3/2< \frac{b}{\bar{c}} < -1$  & $ { \underline{(\ell= 1, m \geqslant 1 )}}$ and $(\ell \geqslant 3 , m \geqslant 2 \ell -3)$     \\ [0.5ex] \hline
      $\frac{b}{\bar{c}} =-3/2$  & $ { \underline{(\ell= 1, m \geqslant 1 )}}$ and $(\ell \geqslant 4 , m \geqslant 2 \ell -3)$     \\ [0.5ex] \hline
      $-2< \frac{b}{\bar{c}} < -3/2$  & ${\underline{ (\ell= 1, m \geqslant 1 )}}$ and $(\ell \geqslant 4 , m \geqslant 2 \ell -4)$     \\  [0.5ex] \hline
     $ -(p+1)/2 < \frac{b}{\bar{c}} < -p/2$, $p \in \mathbb{N}^*$  & ${ \underline{ (\ell= 1, m \geqslant 1 )}}$ and $(\ell \geqslant p+1  , m \geqslant 2 \ell -(p+1))$ \\ [0.5ex] \hline
  \end{tabular}
  \caption{Classification of the boundary bound states (boundary strings)
  as a function of the boundary coupling $b$. Here $m \geqslant 1$ (denoted $m^0$ in the text),
  is the number of particles in the string (a.k.a. the string multiplicity). The 
 underlined strings are the extra boundary strings appearing in the phase $b < -1/2$ and are responsible for the Gaussian fluctuations at large time in that regime.}\label{tab:stability1}
\end{table}
\end{center}

In the Table~\ref{tab:stability1} the constraint $1 \leqslant \ell \leqslant m$ is of course always obeyed, but not all
values of $\ell$ in that interval are always possible, some may be missing depending on
the value of $b/\bar c$ (for instance $\ell=2$ is always missing for $b/\bar c \leqslant - 1$).

\begin{remark}
Let us emphasize that other solutions can be obtained by changing the sign of one or more rapidities inside the string state. Such a solution however correspond to the same quantum state and by restricting ourselves to states of the form \eqref{EqT:BoundaryStringRapidities} we have defined a family of representant of the possible states.
\end{remark}

Physically, these states correspond to clusters of particles that are bound to the wall (`boundary strings') and confined around $x=0$. Here we give 4 examples of wave-functions.
\begin{itemize}
\item
$m^0=n=1$, $\ell=1$: $\Psi_\mu(x)  = i e^{b x}$. For a single boson $n=1$ it is clear that
there will be a one body bound state only for $b<0$, which is when this state becomes 
normalizable and exists in agreement with the table.
\item
$m^0=n=2$, $\ell=1$: $\Psi_\mu(x_1 \leqslant x_2)=-\frac{2 (b-\bar{c}) e^{b (x_1+x_2)-\bar{c} x_2}}{b}$.
This state exists for any $b<\frac{1}{2}$. Hence, the remarquable feature is that already
for $n=2$ there can be a two-boson bound state even for $0<b<\frac{1}{2}$, i.e. for a repulsive wall. 
Note that it decays as $\sim e^{- (\bar c-b) x}$ if only one particle (the second for $x_2=x$) is far from the wall
and as $\sim e^{- (\bar c-2b) x}$ if both are far from the wall (for $x_1=x_2=x$).

\item 
$m^0=n=2$, $\ell=2$: For $n=2$ there is another possible value for $\ell$, which is $\ell=2$ (we recall the constraint
$1 \leqslant \ell \leqslant m$). This two-boson bound state is more complicated that the $\ell=1$ state, and reads
\bea
&& \!\!\!\!\!\!\!\!\!\!\!\!\!\!\!\!\!\!\!\!  \Psi_\mu(x_1 \leqslant x_2) = \\
&& \!\!\!\!\!\!\!\!\!\!\!\!\!\!\!\!\!\!\!\!\!\!\! -\frac{2 e^{-(b+\bar{c}) (x_1+x_2)}}{b (2 b+\bar{c})} 
 \bigg(b (2 b + \bar c)  e^{2 b (x_1+x_2)+\bar{c} (2 x_1+x_2)}+ (b+\bar{c}) \bar c
e^{x_1 (2 b+\bar{c})} +b \bar{c} e^{x_2 (2 b+\bar{c})} \bigg) \nn
\eea

Note that this eigenstate exists only in the restricted range $-1 < b/\bar c < -1/2$. It decays as $\sim e^{- (\bar c+b) x}$ if only one particle (the second for $x_2=x$) is far from the wall
and as $\sim e^{(\bar c+2b) x}$ if both are far from the wall (for $x_1=x_2=x$). Each of 
these two decay rates precisely vanish at one endpoint of the stability range.
 \item
 $m^0=n$, $\ell=1$: $\Psi_\mu(x_1<x_2<\cdots x_n)= i^n n! \prod_{k=1}^{n-1} (1- \frac{k \bar c}{b}) e^{b \sum_i x_i - \bar c \sum_i (i-1) x_i }$. From the Table~\ref{tab:stability1} this eigenstate always exists for
 $b<0$ and exists for 
$m \geqslant 2 + {\rm E}(\frac{2 b}{\bar c})$ for $b>0$. Note that in the range of
existence, the formal vanishing of the prefactor for some integer values of $b/\bar c$ 
is compensated by the norm. It turns out that this state is the 
ground state for some range of parameters, as discussed below.

\end{itemize}

It can be checked that the states in Table~\ref{tab:stability1} are normalizable iff the condition of existence on $(\ell,m^0)$ \eqref{EqT:Existence} are satisfied. Although it is clear that these states correspond to states that are bound to the wall, we do not have a full physical interpretation of them. In particular the physical interpretation of the parameter $\ell$ is lacking. We do not have a satisfying understanding of the remarkable fact that some states can exist even when $b>0$ (repulsive wall). At a very qualitative level, by examining
the alternative form \eqref{LL2} for the Hamiltonian (see discussion there), one could 
argue that the "image" interaction term $- 2 \bar c \sum_{1 \leqslant  i<j \leqslant n}  \delta(x_i+x_j)$ 
produces, when two particles meet at the wall, an effective "additional" attraction to the wall
(an effective $-\delta(x_i)$ term). Finally, the examples studied above
suggest some insight on the domain of stability, but which remains still very partial.
The wavefunction \eqref{EqT:wave} contains terms (many with vanishing coefficients)
of the form $\exp(i \sum_{a=1}^{m^0} \epsilon_a \lambda^{0a} x_{P_a})$
where $\lambda^{0a}$ is given in \eqref{EqT:BoundaryStringRapidities}
(from now on we consider infinite $L$ and set the deviations $\delta=0$).
Consider the decay when all particles are moved to $x>0$, $x_a=x$.
We note that the term where all $\epsilon_a=-1$ leads to a term 
$\exp( - x \frac{m^0}{2} (m^0+1-2 b - 2 \ell))$
whose condition for decay is identical to the upper bound in (both lines of)
Eq. \eqref{EqT:Existence}. We did not find such a simple interpretation for
the other condition $2 b + \ell \bar c>0$ in \eqref{EqT:Existence}.
 
 It is interesting to know, as a function of $b$, how many boundary string states exist with a given 
string multiplicity $m^0$, which we call $N(b,m^0)$. The formula, obtained from Table~\ref{tab:stability1}, is given in the Appendix
\ref{app:number}.


It is useful to introduce the center of mass of the boundary string 
\bea \label{EqT:BoundaryStringCenterOfMass}
 k_{\ell,m^0} := \frac{1}{m^0} \sum_{a=1}^{m^0} \lambda_{0a} = i \left[ b + \bar c \left(\ell-\frac{1}{2}(m^0+1)\right)\right] \, .
\eea
We will see that boundary strings $(\ell,m^0)$ formally share similarities with usual strings with $(k_j,m_j) = (k_{\ell,m^0} , m^0)$. In particular the energy eigenvalue of a boundary string eigenstate is obtained as
\bea \label{EqT:EnergyBoundary}
E_{\ell,m^0} && = \sum_{a=1}^{m^0} (\lambda^{a0})^2 = E(m^0,k_{\ell,m^0}) \nn \\
&& = -m^0 \left[b + \bar c \left(\ell-\frac{1}{2}(m^0+1)\right)\right]^2 - \frac{\bar{c}^2}{12} m^0 \left((m^0)^2-1 \right)
\eea
with $E(m,k)$ as in \eqref{EqT:energy}.

\subsubsection{General case: boundary and bulk strings.} \label{Sec:LL:GeneralState} For any finite $b$, we conjecture from the above study that a general $n$-particles eigenstates can be obtained, in the $L \to \infty$ limit, by partitioning the $n$ particles into $n_s+n_s^0$ clusters of particles. The first $n_s$ clusters of particles are bulk strings with multiplicities $m_j$, $j=1,\cdots,n_s$, string momenta $k_j$ and rapidities inside the string given by \eqref{EqT:UsualString} (setting $\delta$ terms to zero). The remaining $n_s^0$ clusters of particles are boundary strings with `$\ell$-numbers' and multiplicities $(\ell_j,m_j^0)$, and rapidities inside the string as given by \eqref{EqT:BoundaryStringRapidities} (setting $\delta$ terms to zero). The fact that several boundary strings can coexist in the same state is non trivial
\footnote{It leads formally to coinciding rapidities
$\lambda_{\alpha}=\lambda_{\beta}$ for $\alpha \neq \beta$ which is forbidden for an eigenstate at finite $L$.
This restriction may not apply however in the $L \to +\infty$ limit, if the difference $\lambda_{\alpha}-\lambda_{\beta}$ vanishes in that limit. We have not checked in details that this mechanism,
known in several other cases 
\cite{we-flatlong,HagemansCaux}, occurs here, but we are led to conjecture
that it does, from consistency considerations in the calculations of
Section \ref{sec:KPZfullcomp}.}
The values of the different integers must now satisfy $(n_s,n_s^0) \in [0,n]^2$, $(m_j,m_j^0) \in [0,n]^2$, with $ n = \sum_{j=1}^{n_s} m_j +\sum_{j=1}^{n_s^0} m_j^0 \geqslant 1$ (only one of the two sums can be empty) and the integers $\ell_j \in [1,m_j]$ such that the condition of stability of the boundary string \eqref{EqT:Existence} is satisfied. The energy of such a state is $E_\mu = \sum_{j=1}^{n_s} E(m_j,k_j) +\sum_{j=1}^{n_s^0} E(m_j^0,k_{\ell_j,m_j^0}) $ with $E(m,k)$ as in \eqref{EqT:energy}.

\begin{remark}
We should stress that we have no proof that this set of states is complete in the sense that it provides a complete basis of the Hilbert space on which acts the LL Hamiltonian $H_n$. This will be our working hypothesis in the following. Our results on the KPZ/DP problem in the next Section, which are an application of this study, are quite consistent with this
hypothesis. 
\end{remark} 

\subsection{Norm of string states}\label{sec:normLLsec}

The calculation of the norm of string states starting from the finite size formula \eqref{EqT:normGaudin} is a tedious exercice that requires a careful evaluation of many singular terms that are regularized using the string deviations $\delta$, in the same way as for usual strings in the attractive Lieb-Liniger model \cite{cc-07}. We report the details of the derivation in Appendix \ref{app:normGaudin} and only give here the result: given a state characterized by the set of integers $n_s,n_s^0,m_j,m_j^0,\ell_j$ and bulk strings momentas $k_j$ as in Sec.~\ref{Sec:LL:GeneralState} we obtain its norm as
\bea \label{EqT:NormFinal}
\!\!\!\!\!\!\!\!\!\!\!\!\!\!\!\!\!\!\!\! \frac{1}{|| \Psi_\mu ||^2} = && \frac{4^n}{n!} \frac{1}{L^{n_s}}  \prod_{j=1}^{n_s^0} S_{\ell_j,m_j^0}^{(b)} H_{\ell_j,m_j^0}^{(b)} \prod_{i=1}^{n_s} S_{k_i,m_i} H_{k_i,m_i}   \\ \times
&&   \prod_{1\leqslant i < j \leqslant n_s^0}  D_{k_{\ell_i,m_i^0} ,m_i^0 ;k_{\ell_j,m_j^0} ,m_j^0} \prod_{1\leqslant i < j \leqslant n_s}  D_{k_i,m_i;k_j,m_j}  \prod_{i=1}^{n_s^0} \prod_{j=1}^{n_s}  D_{k_{\ell_i,m_i^0} ,m_i^0 ;k_j,m_j}  \nn
\eea
where we have introduced the notations 
\bea \label{EqT:defDSK}
&& D_{k_i,m_i;k_j,m_j} = \frac{4 (k_i - k_j)^2 + \bar{c}^2 (m_i - m_j)^2}{4 (k_i - k_j)^2 + \bar{c}^2 (m_i + m_j)^2} \frac{4 (k_i + k_j)^2 + \bar{c}^2 (m_i - m_j)^2}{4 (k_i + k_j)^2 + \bar{c}^2 (m_i + m_j)^2} \nn \\
&& S_{k,m} =  \frac{2 k \left(\frac{i k}{\bar{c}}-\frac{m}{2}\right)_m }{(2 k+i \bar{c} m) \left(\frac{i k}{\bar{c}}-\frac{m}{2}+\frac{1}{2}\right)_m} \frac{\bar{c}^{m-1}}{2 m^2}  \nn \\
&& H_{k,m} = \frac{1}{\left(-\frac{\bar{c}^2}{b^2}\right)^m \left(\frac{-2 b+\bar{c}+2 i k-\bar{c} m}{2 \bar{c}}\right)_m \left(\frac{2 b+\bar{c}+2 i k-\bar{c} m}{2 \bar{c}}\right)_m}
\eea
Note that $S_{k,m},H_{k,m}$ are even in $k$ for integer $m$ (see below
for equivalent expressions where it is more apparent).
We also introduced 
\bea \label{SH}
\!\!\! \!\!\! \!\!\! && S_{\ell,m^0}^{(b)} = 2 m^0 S_{k_{\ell,m^0},m^0}  \\
\!\!\! \!\!\! \!\!\! && H_{\ell,m^0}^{(b)} = (-1)^{m^0} i {\rm res}_{k=k_{\ell,m^0}} H_{k,m^0}  = \frac{\bar c (-1)^{\ell-1}  (b/\bar{c})^{2m^0}}{(1-\ell-2(b/\bar{c}))_{m^0} (\ell-1)! 
(m^0-\ell)!} \, , \nn
\eea
and we recall the definition \eqref{EqT:BoundaryStringCenterOfMass} of the boundary string center of mass $k_{\ell,m}$. Note that the norm of a state scales as $L^{n_s}$ as a signature of the fact that the bulk strings are delocalized and almost behave as free particles. On the other hand the norm of a pure boundary string does not scale with $L$, showing that these states are localized in space. The formula
 \eqref{EqT:NormFinal}, presented here as a conjecture, has been checked numerically for states with
 a small number of particles.
 
In the hard wall limit $b \to +\infty$ studied in \cite{GueudrePLD}, one has
$\lim_{b \to +\infty} H_{k,m}=1$ and the boundary strings do not exist. The 
above formula \eqref{EqT:NormFinal} reproduces formula (9)-(10) of \cite{GueudrePLD},
correcting for the misprint there, i.e. replacing there $S_{k_i,m_i} \to S_{k_i,m_i}^{-1}$ and 
$D_{k_i,m_i,k_j,m_j} \to D_{k_i,m_i,k_j,m_j}^{-1}$ in all three lines of (9-10). With these
corrections, the definition of the factor $S_{k,m}$ there 
equals $2^{2m-1}$ the factor $S_{k,m}$ here, and the $D$ factors are the same.
Note that all other definitions there, such as the wavefunction \eqref{EqT:wave}, are the same as here.

\begin{figure}
\centering
\includegraphics[width=12cm]{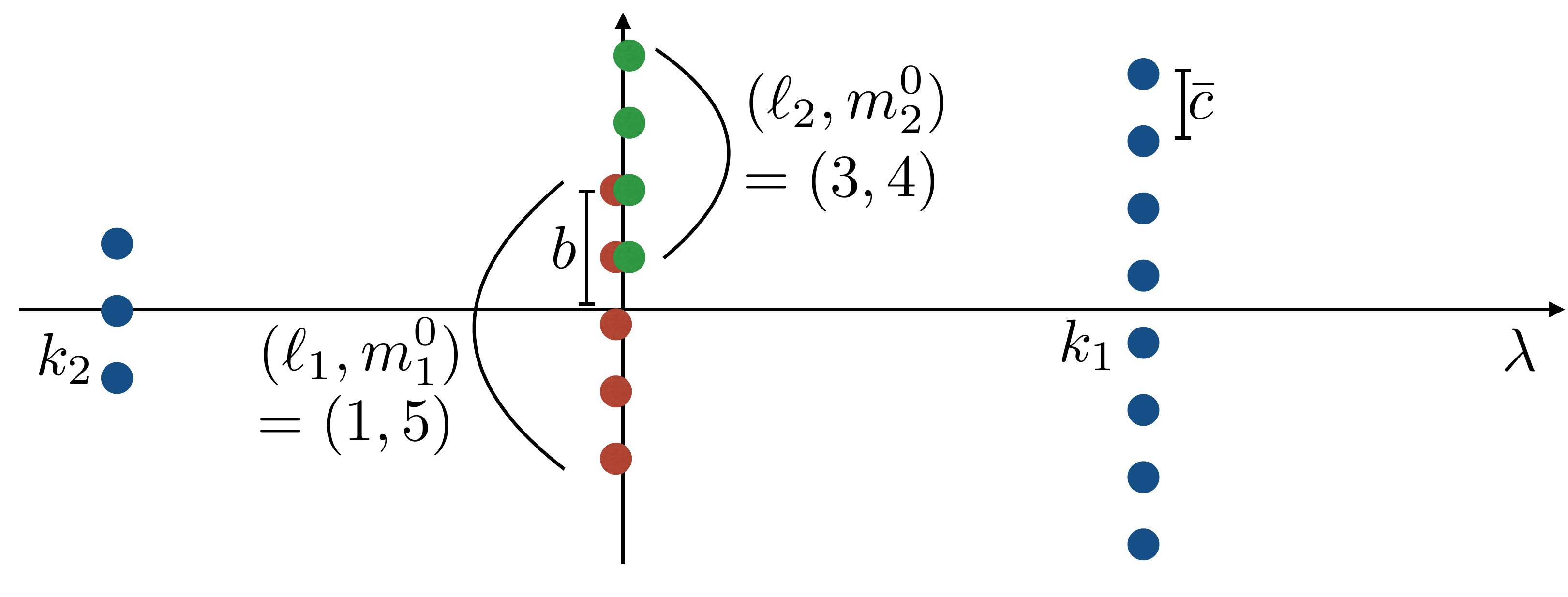}
\caption{A general state of $n$ particles is obtained by arranging the rapidites inside an arbitrary number of bulk and boundary strings. Here $n=20$ and $n_s^0=n_s = 2$; $m_1=8$ and $k_1>0$; $m_2=3$ and $k_2<0$; $(\ell_1,m_1^0)=(1,5)$ (red dots) and $(\ell_2,m_2^0) = (3,4)$ (orange dots).}
\label{fig:string3}
\end{figure}

\subsection{Ground state of the attractive Lieb-Liniger on the half-line}\label{sec:GSLLb}

\begin{figure}
\centering
\includegraphics[width=10cm]{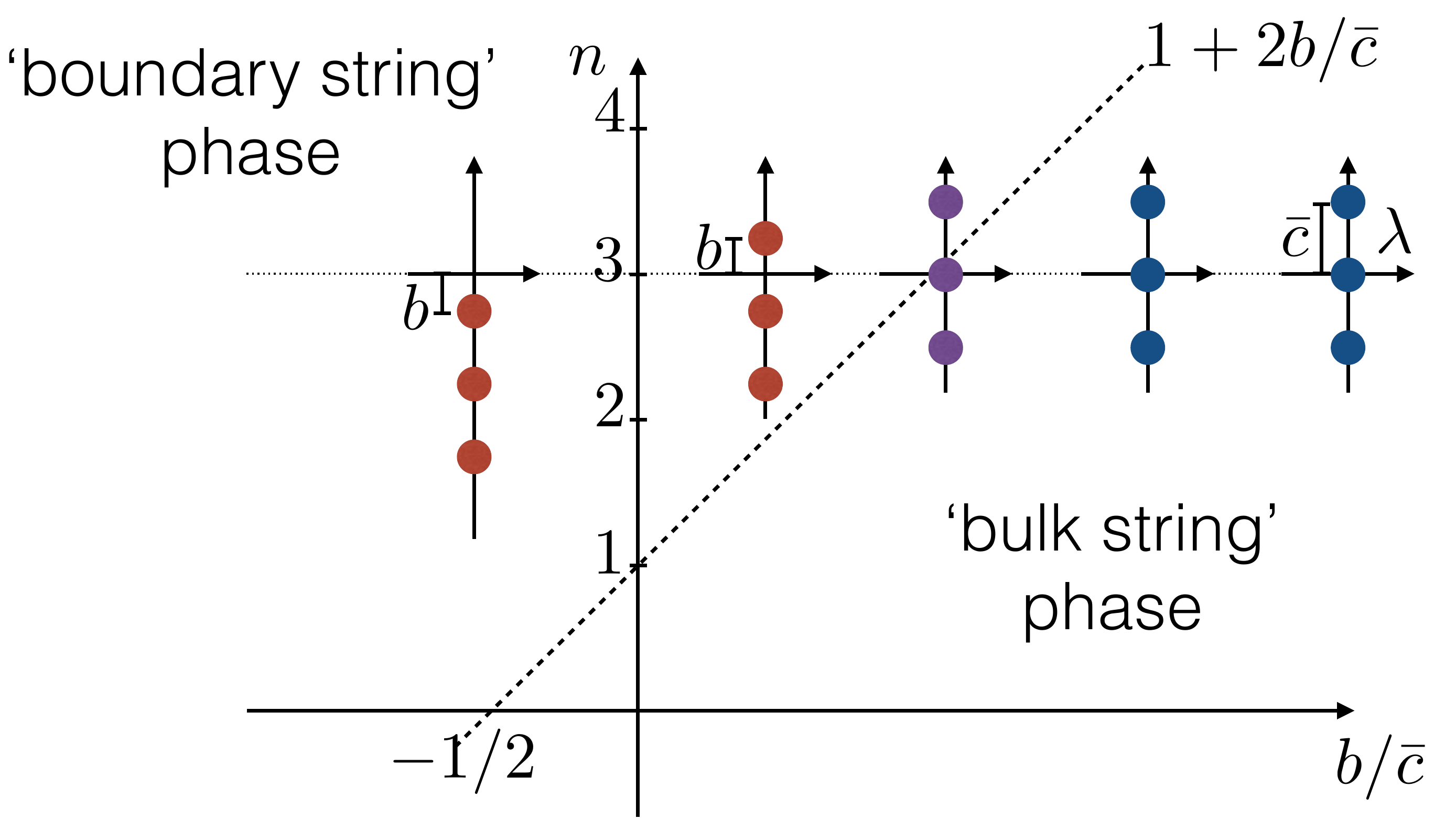}
\caption{Ground state phase transition in the Lieb Liniger model \eqref{LL} with boundary 
as a function of the boundary parameter $b$ in \eqref{bc1}. For $n$ fixed, the ground state exhibits a continuous phase transition on the dotted line $n=1 + \frac{2 b}{\bar c}$ 
from a bulk dominated phase at large $b$ (the ground state is given by a single bulk string) to a boundary dominated phase at small $b$ (ground state given by a single boundary string). The transition in the complex plane for
the corresponding rapidities inside the string is sketched for $n=3$.}
\label{fig:QuantumGroundState}
\end{figure}

We now discuss the properties of the ground state for a system of $n$ particles as a functions of $n$ and $b$. In the absence of states bound to the wall ($b = +\infty$), reminiscent of the usual attractive LL model on the full space, it is well known that the ground state is obtained by putting all particles in a single string state with string momenta $k=0$. We recall that the energy for a string of $n$ particles with string momenta $k$ is (see \eqref{EqT:energy})
\bea
E(n,k) = n k^2 - \frac{\bar{c}^2}{12} n(n^2-1) \, .
\eea
And the ground state energy is thus in the $b = +\infty$ case $E_{{\rm GS}} = - \frac{\bar{c}^2}{12} n(n^2-1)$. In the finite $b$ case one can hope to find states with lower energy since the energy of the $(\ell,n)$ boundary string is $E(n,k_{\ell,n})$ with $k_{\ell,n}$ {\it imaginary} and given by \eqref{EqT:BoundaryStringCenterOfMass}: we recall that
\bea
E(n,k_{\ell,n}) = - n \left[b + \bar c \left(\ell-\frac{1}{2}(n+1)\right)\right]^2 - \frac{\bar{c}^2}{12} n (n^2-1) \, .
\eea
Note that as a function of $\ell$, $E(n,k_{\ell,n})$ is a parabola which reaches its maximum for $\ell = \ell_{{\rm max}} := -\frac{b}{\bar{c}} +\frac{1}{2}(n+1)$. Since this value coincides with the upper bound on $\ell$ that is set by the condition of existence of the boundary bound state \eqref{EqT:Existence}, we can conclude that the boundary bound state with the minimum energy is obtained by taking $\ell=1$ which indeed always exist if any boundary bound state does. The energy of this state is 
\be \label{groundstateenergy} 
E_b(n):= E(n,k_{1,n}) = - n \left(b + \frac{\bar c}{2} (1-n)\right)^2 - \frac{\bar{c}^2}{12} n (n^2-1)
\ee
and we can conclude that this state is the ground state whenever it exists, that is when $n > 1 + \frac{2b}{\bar{c}}$. To summarize we have obtained that
\begin{itemize}
	\item
	For $n\leqslant 1+2b/\bar c$ the ground state is a state made of a single bulk string with vanishing momenta. The ground state energy is $E_{{\rm GS}} = E_b(n)=- \frac{\bar{c}^2}{12} n (n^2-1)$. The ground state wavefunction 
in the bulk (i.e. for all points far from the boundary\footnote{near the boundary
it takes a more complicated, $k$-dependent, form where $k$ is the (small) momentum.}) looks like the ground state of the full space problem
\be \label{45} 
\Psi_{{\rm GS}}(x_1,\cdots,x_n) \sim n! \exp\left(-\frac{\bar c}{2} \sum_{1\leq \alpha < \beta \leqslant n} |x_\alpha - x_\beta| \right) 
\ee 
\item
For $n>1+2b/\bar c$ the ground state is a state made of a single boundary string with $\ell=1$. The ground state energy is $E_{{\rm GS}} = - n \left(b + \frac{\bar c}{2} (1-n)\right)^2 - \frac{\bar{c}^2}{12} n (n^2-1)$. 
Formula 
\eqref{EqT:NormFinal}-\eqref{EqT:defDSK}-\eqref{SH} for the norm
gives
\bea
&& || \Psi_\mu ||^2 = 
 \frac{n!^2}{4^n} b^{-2n} \bar c^n
\frac{2 b + \bar c}{2 b + \bar c(1-n)} \frac{(-b/\bar{c})_{n} 
(-2b/\bar{c})_{n}}{ (- \frac{1}{2} -b/\bar{c})_{n}}.
\eea 
Hence the normalized eigenstate is
\begin{align}
  \Psi_{{\rm GS}}&(x_1 \leqslant \cdots \leqslant x_n) =  (-1)^{n-1}  (2 i)^n b \bar c^{\frac{n}{2}-1} 
\sqrt{\frac{2 b + \bar c(1-n)}{2 b + \bar c} \frac
{ (- \frac{1}{2} -\frac{b}{\bar{c}})_{n}}{(-\frac{b}{\bar{c}})_{n} (-\frac{2b}{\bar{c}})_{n}}}
\nn \\& \hspace*{3cm }\times (1-\frac{b}{\bar{c}})_{n-1} e^{b \sum_i x_i - \bar c \sum_i (i-1) x_i }. \nn
\end{align}
\end{itemize}
The transition occurs for $n = 1 + \frac{2b}{\bar{c}}$. Exactly at the transition the two states are formally identical as their rapidities are the same (nondegenerate ground state) 
i.e. \eqref{EqT:BoundaryStringRapidities} and \eqref{EqT:UsualString} coincide,
$\lambda^{0a}= i b + i \bar c(1-a) = \frac{i \bar c}{2} (n+1-2 a)$,
but the state should be considered as a bulk string ground state as it is delocalized in the full volume. The ground state energy is continuous across the transition. A cartoon of the transition is shown in Fig.~\ref{fig:QuantumGroundState}.

\begin{remark}
One can explore the possibility that the ground state is
composed of several strings. For bulk strings it is well known that splitting
them only leads to excited states. Let us consider the state with two boundary strings
of particle content $m_1^0=m$ and $m_2^0=n-m$. For each of them the same arguments as above, replacing $n$ by $m_j^0$ shows that $\ell=1$ is the best choice. We now note that 
\be
E_b(m) + E_b(n-m) - E_b(n) = \bar c m (n-m) (\bar c (n-1)-2 b) 
\ee
is always strictly positive for $1 \leqslant m  \leqslant n-1$ and in the domain $\bar c n  >2 b+ \bar c$. 
For $b<0$ this is always true for $n>1$. For $b>0$, since we have assumed that $\bar c m_j^0 >2 b + \bar c$ (the condition of stability for each string) it implies $\bar c n > 2 (2 b+ \bar c) > (2 b + \bar c)$. 
Hence it is never advantageous to split a boundary string into two boundary strings
and the ground state is thus composed of a single boundary string in the region $n>1+2b/\bar c$
as stated above.
\end{remark}

\subsection{Decomposition of the identity in terms of Lieb-Liniger eigenstates}

We conclude this section by writing explicilty the decomposition of the identity that follows from our classification of states as a function of $b$. Let us first write the formula and then comment it. On the space of symmetric functions of $n$ variables $(x_1,\cdots,x_n) \in [0,+\infty[^n$ with the appropriate boundary condition \eqref{EqT:Spectral2} at $x=0$ the identity can be written as
\bea \label{EqT:decompositionidentity}
\mathbb{I} =&&  \sum_{n_s \geqslant 0} \sum_{n_s^0 \geqslant 0} \sum_{m_1, \cdots , m_{n_s} \geqslant 1} \sum_{(\ell_1,m_1^0), \cdots , (\ell_{n_s^0} , m_{n_s^0})}^{(b)} \frac{1}{n_s!} \frac{1}{n_s^0!} \delta\left(n- \sum_{i=1}^{n_s} m_i - \sum_{i=1}^{n_s^0}m_i^0 \right) \nn \\ \times
&& \prod_{j=1}^{n_s} \int_{0}^{\infty} \frac{L m_j \rmd k_j}{\pi} \frac{|\Psi_\mu \rangle \langle \Psi_\mu |}{||\Psi_\mu ||^2}
\eea
where
\begin{itemize}
	\item
	$|\Psi_\mu\rangle$ denotes the state with rapidites arranged in boundary and bulk strings according to the numbers $n_s,n_s^0,m_j,m_j^0,\ell_{j}$ and $k_j$ as explained before.
	\item
	The sum $\sum_{(\ell,m^0)}^{(b)}$ denotes the sum over the possible integers $(\ell,m^0)$ parametrizing stable boundary strings as a function of $(b)$ (see \eqref{tab:stability1} for the domain of stability).
	\item
	The factorial factors $\frac{1}{n_s!} \frac{1}{n_s^0!}$ avoid multiple countings of the same state.
	\item
	The phase space factor $ \frac{L m_j \rmd k_j}{\pi}$ is standard and can be determined by studying the reduced Bethe equations satisfied by the string momenta $k_j$ \cite{cc-07}. Note two differences compared to \cite{cc-07} here: the quantization is $ \frac{L m_j \rmd k_j}{\pi}$ and not $ \frac{L m_j \rmd k_j}{2 \pi}$ due to the additional factor of $2$ inside the exponential on the left-hand side of the Bethe equations \eqref{EqT:BetheEq1}; the integral over $k_j$ only runs from $0$ to $+\infty$ since states with string momenta $k_j$ and string momenta $-k_j$ are identical.
\end{itemize}

\begin{remark}
The formula for the decomposition of the identity \eqref{EqT:decompositionidentity} can be modified using the symmetries of the Bethe states. In particular $ \int_{0}^{\infty} \frac{L m_j \rmd k_j}{\pi}$ can be transformed into $ \int_\mathbb{R} \frac{L m_j \rmd k_j}{2\pi}$. This will be useful in the section on the application to the directed problem in order to obtain formulas with the desired symmetries made explicit.
\end{remark}

\section{Application to the KPZ equation/directed polymer problem on the half-plane}\label{sec:KPZfullcomp}

In this section we use our study of the Lieb Liniger model on a half-line to study the continuum directed polymer/KPZ equation on a half-space as introduced in Sec.~\ref{Sec:Models:KPZDP}. We start by giving our result for the integer moments of the partition sum of the DP for fixed endpoints near the wall. Given these formulas, we already show that the existence, and some properties, of the transition at $b=-1/2$ 
as the interaction parameter with the wall $b$ is varied, can be obtained from the contribution of the
quantum ground state of the Lieb Liniger model. These results confirm earlier results by Kardar \cite{KardarTransition} and put them on a firmer basis. We then use our expressions for the
integer moments (which are valid at all times) to obtain the exact statistics of the KPZ height-field/directed polymer free-energy, at large time/large polymer length $t$. We show that the fluctuations of the appropriately rescaled variable changes from GSE-TW type fluctuations ($b>-1/2$) of width $t^{1/3}$ to Gaussian fluctuations ($b<-1/2$) of width $t^{1/2}$, with at the transition point $b=-1/2$ GOE-TW type fluctuations of width $t^{1/3}$. We recall that, here and throughout this section, the DP model defined in Sec.~\ref{Sec:Models:KPZDP} is associated, as in Sec.~\ref{Sec:Models:KPZandLL}, to the LL model studied in the previous section where the parameters $\bar c$ has been set to $1$ by a choice of units.

\subsection{Moments of the DP partition sum for fixed endpoints near the wall}
\label{sec:momentsZ} 

\subsubsection{Starting formula for the moments}

From now on we focus on the partition sum of the directed polymer $Z_\eta(x,t|y,0)$ introduced in Sec.~\ref{Sec:Models:KPZDP} for fixed endpoints at the wall $x=y=0$ (droplet initial conditions in the KPZ framework). We use in the following $Z(t)$ as a shorthand, $Z(t)=Z_\eta(0,t|0,0)$. We start from the formula \eqref{EqT:ZnLL1} giving the moments $\overline{Z(t)^n}$ in terms of the eigenstates of the LL model and now make it explicit using the results of the previous section. Using $\Psi_\mu(0,\cdots,0) = \frac{n!}{b^n} \prod_{\alpha=1}^{n} \lambda_\alpha$, and decomposing an arbitrary eigenstate in terms of boundary and bulk strings as in Sec.~\ref{Sec:LL:GeneralState}, it is a simple exercice to see that we can write
\bea
|\prod_{\alpha=1}^{n} \lambda_\alpha|^2 = \prod_{i=1}^{n_s} a_2(k_j,m_j) \prod_{i=1}^{n_s^0} (-1)^{m_i^0}a_2(k_{\ell_i,m_i^0} , m_i^0)
\eea 
where we have introduced
\bea \label{EqT:defa2}
\!\!\!\!\!\!\!\!\!\!\!\!\!\!\!\!\!\!\!\!\!\!\!\!\!\!\! a_2(k,m) :=  (i k + \frac{1-m}{2})_m (- i k + \frac{1-m}{2})_m =
\frac{\Gamma(\frac{1+m}{2} + i k) \Gamma(\frac{1+m}{2} - i k)}{\Gamma(\frac{1-m}{2} + i k)\Gamma(\frac{1-m}{2} - i k)} \, .
\eea
and used that $(-i k + \frac{1-m}{2})_m= (-1)^m (i k + \frac{1-m}{2})_m$ for integer $m$.
Note that $a_2(k,m)$ is explicitly even in $k$ and that for $m$ integer the product 
$S_{k,m} a_2(k,m)$ is an even polynomial in $k$. Combining this with the decomposition of the identity \eqref{EqT:decompositionidentity} and the formula for the norm of string states \eqref{EqT:NormFinal} we get 
\begin{align} \label{EqT:StartingZn}
\overline{Z(t)^n} =&  b^{-2n} n! 4^n  \sum_{n_s\geq 0} \sum_{n_s^0 \geqslant 0} \sum_{m_1 , \cdots m_{n_s}  \geqslant 1} \sum_{(\ell_1,m^0_1), \cdots , (\ell_{n_s^0},m^0_{n_s})}^{(b)}  e^{-t \left[ \sum_{j=1}^{n_s^0}  E(m_j^0,k_{\ell_j,m_j^0}) +  \sum_{j=1}^{n_s}  E(m_j,k_j) \right]}  \nn \\ \times
& \delta\left(\sum_{j=1}^{n_s} m_j +  \sum_{j=1}^{n_s^0} m_j^0 - n \right)   \frac{1}{n_s!}\frac{1}{n_s^0!}    \nn \\  \times
& \prod_{j=1}^{n_s^0}  2 m_j^0 S_{k_{\ell_j,m_j^0} ,m_j^0} a_2(k_{\ell_j,m_j^0} ,m_j^0) i {\rm res}_{k=k_{\ell_j,m_j^0}} H_{k,m_j^0}  \nn \\ \times
& \prod_{j=1}^{n_s}  \int_0^{\infty} \frac{d k_j}{\pi} m_j S_{k_j, m_j} H_{k_j, m_j}  a_2(k_j,m_j) \nn \\ \times
& \prod_{1\leqslant i<j \leqslant n_s^0} D_{k_{\ell_i,m_i^0},m_i^0,k_{\ell_j,m_j^0},m_j^0}   \prod_{1\leq i<j \leqslant n_s} D_{k_i,m_i,k_j,m_j}  \prod_{i=1}^{n_s^0} \prod_{j=1}^{n_s} D_{k_{\ell_i,m_i^0},m_i^0,k_j,m_j} \,.
\end{align}

\subsubsection{Symmetrization}

We now obtain an equivalent but more symmetric formula. We introduce for simplicity the following function
\bea \label{EqT:defBkm}
\fl && B_{k,m} = 2m^2 4^m S_{k,m} H_{k,m} a_2(k,m)  \\
\fl && =\frac{b^{2m} 2 k }{\pi} \sinh(2 \pi k)  \Gamma(m+ 2 i k) \Gamma(m- 2 i k) 
 \frac{\Gamma\left(b + \frac{1-m}{2} - i k \right)  \Gamma\left(b + \frac{1-m}{2} + i k\right)}{ 
\Gamma\left(b + \frac{1+m}{2} - i k \right)  \Gamma\left(b + \frac{1+m}{2} + i k\right)} \nonumber
\eea
that enters into \eqref{EqT:StartingZn}.
We recall the definitions \eqref{EqT:defDSK} of the functions $S_{k,m}$ and $H_{k,m}$ and \eqref{EqT:defa2} of $a_2(k,m)$. It is easily checked that $B_{k_j,m_j}$  is even in $k_j$, as are the other factors that contain $k_j$ in \eqref{EqT:StartingZn}, i.e.
$D_{k_i,m_i,-k_j,m_j}=D_{k_i,m_i,k_j,m_j}$. The integration on $k_j >0$ can be extended to an integration on $k_j \in \JR$, adding an addition factor $1/2$ in the measure. To make the contribution of boundary strings more symmetric, we will use the following notational trick. For a function $f(k)$ which has a simple pole 
at $k = k_{\ell,m^0}$ we write
\be \label{trick}
A(k_{\ell,m^0}) \, {\rm res}_{k = k_{\ell,m^0}} f(k) 
=  \int_{k^0 \in \JR} \rmd k^0 \delta(k^0-k_{\ell,m^0}) (k^0-k_{\ell,m^0}) f(k) A(k)  \, .
\ee 
which we will apply to $f(k)=H_{k,m^0}$. The factor $A(k)$ will contain 
all the other factors $S,a,D$ in \eqref{EqT:StartingZn}. For generic (non integer) values of $b$ they do not
have poles at $k_{\ell,m^0}$ hence they can be entered into the residue
${\rm res}_{k = k_{\ell,m^0}} f(k) A(k)$. Even then, the integral in \eqref{trick} 
is formal since $k_{\ell,m^0} \in i \JR$. Note that 
${\rm res}_{k = k_{\ell,m^0}} B_{k,m^0} = 2 (m^0)^2 4^{m^0} S_{k_{\ell,m^0},m^0}
a_2(k_{\ell,m^0},m^0) {\rm res}_{k = k_{\ell,m^0}}  H_{k,m^0}$ since the product
$S a_2$ is a polynomial (see \eqref{limit} below), hence we can indifferently apply the
residue operation to $H$ or to $B$. 

This allows to introduce $n_s^0$ (fictitious) integrals over boundary string momentas $k_j^0$. Using finally the (anti-)symmetry of the residue, ${\rm res}_{k = k_{\ell,m^0}} B_{k,m} = - {\rm res}_{k = -k_{\ell,m^0}} B_{k,m}  $, we rewrite \eqref{EqT:StartingZn} as  
\begin{align}  \label{EqT:SymmetrizedZn} 
\overline{Z(t)^n} =&   n! b^{-2n}  \sum_{n_s\geq 0} \sum_{n_s^0 \geqslant 0} \sum_{m_1 , \cdots m_{n_s}  \geqslant 1}  \sum_{(\ell_1,m^0_1), \cdots , (\ell_{n_s^0},m^0_{n_s})}^{(b)}    \nn \\
& \delta\left(\sum_{j=1}^{n_s} m_j +  \sum_{j=1}^{n_s^0} m_j^0 - n \right)   \frac{1}{n_s!}\frac{1}{n_s^0!}    \nn \\  \times
& \prod_{j=1}^{n_s^0}  \int_\mathbb{R} \frac{\rmd k_j^0}{2\pi}   \frac{B_{k_j^0,m_j^0}}{2m_j^0}  (2i\pi) \left[ (k_j^0-k_{\ell_j,m_j^0}) \delta(k_j^0-k_{\ell_j,m_j^0}) -  (k_j^0+k_{\ell_j,m_j^0}) \delta(k_j^0+k_{\ell_j,m_j^0}) \right]\nn  \\& \times \prod_{j=1}^{n_s}  \int_\mathbb{R} \frac{\rmd  k_j}{(2 \pi)} \frac{B_{k_j,m_j}}{2m_j} \prod_{1\leq i<j \leqslant n_s^0} D_{k_i^0,m_i^0,k_j^0,m_j^0}   \prod_{1\leq i<j \leqslant n_s} D_{k_i,m_i,k_j,m_j}  \prod_{i=1}^{n_s^0} \prod_{j=1}^{n_s} D_{k_i^0,m_i^0,k_j,m_j}\nn\\& 
\times e^{-t \left[ \sum_{j=1}^{n_s^0}  E(m_j^0,k_j^0) +  \sum_{j=1}^{n_s}  E(m_j,k_j) \right]} \, .
\end{align}
where we recall that the integral over $k_j^0$ with the delta function terms simply and only means
to take the corresponding residue in the factor $B_{k_j^0,m_j^0}$. 

Eq. \eqref{EqT:SymmetrizedZn} is our main final exact result for the moments of the DP partition sum, equivalently the exponential moments of the KPZ height at the boundary,
$\overline{e^{n h(0,t)}}$, for droplet initial conditions. We stress that it is valid at all times $t$
and for any $b$. 

\begin{remark}
In the limit $b \to +\infty$, the factor $H_{k,m}$ becomes unity and 
\bea \label{limit} 
&& \! \! \! \! \! \! \! \! \! \! \! \! \! \! \! \! \! \! \! \! \! \! \! \! \! \! \! \! \! \! \lim_{b \to +\infty} B_{k,m} = b_{k,m} := 2 m^2 4^m S_{k,m} a_2(k,m) = \prod_{j=0}^{m-1} (4 k^2 + j^2) \\
&& 
= \frac{2 k}{\pi} \sinh(2 \pi k) \Gamma(m+2 i k) \Gamma(m-2 i k) \nn
\eea
for integer $m$. There are
no boundary strings $n_s^0=0$ and the formula \eqref{EqT:SymmetrizedZn} identifies with Eq. (11)
in \cite{GueudrePLD} (with $D$ the same factor as here, note the misprint in (10) $D$ there should be $D^{-1}$, see discussion at the end of section \ref{sec:normLLsec}). Note that $Z$ in the l.h.s. of Eq. (11) of 
\cite{GueudrePLD} should be identified as $Z= b^2 Z(t)$,
which has a finite limit for $b=+\infty$, also equal to $Z=\lim_{x \to 0} Z(x,t)/x^2$ (Eq. (8) there). 
\end{remark}

\begin{remark}
In the domain $b + \frac{1}{2} > n/2$ there are no boundary strings and formula 
\eqref{EqT:SymmetrizedZn} simplifies to the expression given in (38) in Ref. \cite{AlexLD}
(using the duality to the Brownian IC) and in Ref. \cite{PLD1} (via a direct calculation).
\end{remark}

\subsection{Ground state physics and KPZ height function }\label{sec:GSandKPZ}

The ground state phase transition discussed in the Lieb-Liniger context in Section \ref{sec:GSLLb} allows to obtain some properties of the phase transition in the directed polymer context. Here the interesting quantity is the directed polymer quenched free-energy $F(t) =- \log Z(t)$ in our units. We will focus on the 
KPZ height field $h(t)=h(0,t)$, which is equivalent, and simply equal to {\it minus} the free energy
of the DP,  $h(t) = \log Z(t) = -F(t)$. At large time, both grow linearly plus fluctuations
\bea
h(t) \simeq v_\infty(b) \, t + \delta h(t)   \quad , \quad F(t) \simeq - v_\infty(b) \, t - \delta h(t)
\eea 
with $\delta h(t)$ is a random variable of order $\sim t^{\beta(b)}$, and we denote by $v_\infty(b)$ the KPZ growth speed.
Here $\beta=\beta(b)$ is the growth exponent, to be determined below.
In the DP context, $\beta$ is also called $\theta$, the free energy fluctuation exponent.

The average free-energy per-unit length $- v_\infty(b)$ can be obtained from the replica trick in the 
$n=0$ limit as
\bea
v_\infty(b) := - \lim_{t \to \infty} \frac{1}{t}\overline{F(t)} = \lim_{t \to \infty} \lim_{n \to 0} \frac{1}{t} \frac{\overline{Z^n} -1}{n} \, .
\eea
Inverting the limit on $t$ and $n$, one approximates $Z^n$ by $e^{-t E_{{\rm GS}}(n)}$ (i.e. retaining only the ground state contribution for $t$ large). The explicit expression for the ground state energy $E_{{\rm GS}}(n)$ of the LL model in a half-space was given in Section \ref{sec:GSLLb}. 
Performing its analytical continuation to real $n$ in the naive way we can expand around $n =0$ 
and we obtain two cases (see Figure \ref{fig:QuantumGroundState}). 
\begin{itemize}
\item For $1+2b>0$  (recall that $\bar c =1$ in our present unit) the ground state is for $n$ sufficiently close to $0$ obtained by putting all particles in a bulk string with zero momentum and we obtain, expanding $e^{-t E_{{\rm GS}}(n)}$ in $n$
\be
v_\infty(b)  =  \frac{1}{t} \lim_{n\to 0} \frac{ t  \frac{1}{12} n (n^2-1) }{n} = - \frac{1}{12} \quad \text{ for $1+2b >0$.}
\ee
which is the standard value for the full space continuum KPZ equation\footnote{We recall that this negative value comes from the Ito prescription in the continuum SHE with white noise. If e.g. a space-time cutoff on the noise is used instead, an additional, non universal positive constant is added to $v_\infty(b)$ (in both phases, and at most smoothly varying with $b$, so it does not affect the universality of the transition studied here).}.
\item For $1+2b \leqslant 0$, the ground state is always obtained as a boundary string $(\ell=1,n)$ and we 
obtain
\be \label{vinfty} 
\begin{split}
& v_\infty(b)  =  \frac{1}{t} \lim_{n\to 0} \frac{ t  \left( n \left(b + \frac{1}{2}(1-n) \right)^2 + \frac{1}{12} n (n^2-1)  \right)}{n} =  - \frac{1}{12}  + \left(b +\frac{1}{2} \right)^2  \\
&  \text{ for $1+2b\leqslant 0$.}
\end{split}
\ee
hence the free energy per unit length, $- v_\infty(b)$, is lower in this phase. 
\end{itemize}
We can therefore conclude that the DP exhibits a phase transition at $b =-1/2$. This is a binding transition in presence of disorder, from a bulk phase $b >-1/2$ where the polymer rarely comes back near the wall, to a bound phase for $b <-1/2$, where the polymer spends a macroscopic $\mathcal{O}(t)$ fraction of its length near the wall. We thus confirm here the arguments of
Kardar \cite{KardarTransition}, via a detailed analysis of the boundary strings (only the state $\ell=1$ was
considered there) \footnote{In our units, the correspondence holds setting 
the parameters there $\tilde{b} = \frac{1}{2}$, $\lambda=-b$, $\gamma=1$ and $\sigma^2=2$.}.
Note that at this stage we have not specified the initial conditions (for KPZ) i.e. the
endpoints conditions for the polymer. 

\begin{remark}
Note that this most naive replica trick correctly predicts the extensive part of the free energy,
i.e. predicts $v_\infty(b)$ via the linear term in $n$ in $E_{{\rm GS}}(n)$. One could attempt to expand
in powers of $n$ near $n=0$ to predict fluctuations. In the full space problem, it is well known that it 
fails, as the $t\to +\infty$ and $n \to 0$ (or analytical continuation in $n$) do not commute since
the $L=+\infty$ limit has already been taken \cite{bb-00}. Hence we can anticipate that here
in the unbound phase it also fails. This will be discussed again below. 
\end{remark}

\subsection{Rescaled free-energy, KPZ height and definition of the generating function}\label{sec:defgeneratingfunc}

In order to study the statistics of the free-energy $F(t)$ of the DP at large time we decompose 
(for arbitrary time) 
\bea \label{dec1} 
h(t) = - F(t)=  v_\infty(b) t + \left( \frac{t}{4} \right)^{\beta(b)}  \tilde{h} \, ,
\eea
with $v_\infty(b)$ given in the precedent section and $\tilde{h}$ is the centered and scaled 
KPZ height, a random variable. In our units (which
amount to set $\bar c=1$ in the LL model) we thus have 
\begin{equation}
v_\infty(b) = \begin{cases} 
      - \frac{1}{12} & b \geqslant 1/2 \\
     -\frac{1}{12}+(b+\frac{1}{2})^2 & b<-1/2 \, .
   \end{cases}
\end{equation}
As written in \eqref{dec1} the probability distribution function (PDF) of (minus) the rescaled height function $\tilde{h}$ 
depends on $t$, but here we will show that, choosing the growth exponent $\beta(b)$ as
\begin{equation}
\beta(b) = \begin{cases} 
      1/3 & b \geqslant -1/2 \\
      1/2 & b<-1/2 \, ,
   \end{cases}
\end{equation}
then $\tilde{h}$ is an $\mathcal{O}(1)$ random variable whose PDF in the infinite time limit will be computed below. We will show that for $b>-1/2$, $\tilde{h}$ is distributed according to the GSE-TW distribution, for $b=-1/2$, $\tilde{h}$ is distributed according to the GOE-TW distribution and for $b<-1/2$, $\tilde{h}$ is distributed according to a Gaussian distribution.

To compute the distribution of $\tilde{h}$ we first define the following generating function:
\begin{eqnarray}  \label{defg0}
\!\!\!\!\!\!\!\!\!\!\!\!\!\!\!\!\!\!  g_b(s) = \overline{ \exp(- e^{- (t/4)^{\beta(b)} s - t v_\infty(b)} Z(t)) } = 1 + \sum_{n=1}^\infty \frac{(- e^{-  (t/4)^{\beta(b)} s - t v_\infty(b)})^n}{n!} \overline{Z(t)^n} 
\end{eqnarray} 
Let us introduce a unit Gumbel random variable $G$, statistically independent from
$\tilde{h}$, with 
cumulative distribution function (CDF), $\mathbb{P}(G < z) = e^{-e^{-z}}$. 
Then from \eqref{defg0} 
\be
g_b(s) = \mathbb{P}\left(\tilde{h} + \left(\frac{4}{t} \right)^{\beta(b)} G < s \right) 
\ee 
which is an exact formula for all $t$. In the large time limit it simplifies as
\be
 \lim_{t \to \infty} g_b(s) = \lim_{t \to \infty} \mathbb{P}(\tilde{h}  < s) 
\ee
i.e. the CDF of the centered and scaled KPZ height at infinite time
coincides with the large time limit of the generating function. 
Our strategy is thus as follows: we first compute \eqref{defg0} using the expression for  $\overline{Z(t)^n} $  \eqref{EqT:SymmetrizedZn} and show that $g_b(s)$ can be written as a Fredholm 
Pfaffian at all time $t$. We will then perform an asymptotic analysis of these formulas at large time in the different phases, sheding light on the phase transition at the level of the fluctuations of $\tilde{h}$. Anticipating the calculations, we obtain
\begin{itemize}
\item \textbf{Unbound regime:} $b>-1/2$. In the limit $t \to +\infty$ one has
\begin{equation}
\lim_{t \to \infty} \mathbb{P}(\tilde{h}  < s)  = \sqrt{\text{Det}\left( I - P_{  2^{-2/3}  s} K^{\rm GLD} P_{ 2^{-2/3} s} \right)} = F_4(2^{-2/3} s)
\end{equation}
where
\begin{equation}\label{eq:GSEKErnel1}
K^{\rm GLD}(v_i,v_j) = K_{\Ai}(v_i,v_j)  - \frac{ \Ai(  v_i)  }{2} \int_{0}^\infty \rmd y    \Ai(  y + v_j) \, .
\end{equation}
was proved to be the one-dimensional GSE kernel 
\cite{GueudrePLD,krajenbrink2018large}. Here $F_4$ is the CDF of the GSE-TW distribution
(same convention as in \cite{GueudrePLD,AlexLD}). 

\item \textbf{Critical boundary regime:} $b=-1/2$. In the limit $t \to +\infty$ one has
\begin{align}
 \lim_{t \to \infty}\mathbb{P}(\tilde{h}  < s)  
 = \sqrt{\text{Det} \left( I  - P_{ 2^{-2/3}{ s}}K_{\rm GOE} P_{ {  2^{-2/3}  s}} \right)} = F_1(2^{-2/3} s),
 \end{align}
 with the GOE kernel expressed as 
 \begin{equation}
 K_{\rm GOE}(v_i,v_j) = K_{\Ai}(v_i,v_j) + \Ai(v_i) - \Ai(v_i) \int_0^{+\infty} \rmd y \Ai(y + v_j).
 \end{equation}
 and $F_1$ the CDF of the GOE-TW distribution.
 \item \textbf{Attractive boundary regime:} $b<-1/2$.  In the limit $t \to +\infty$ one has
  \begin{equation}
 \tilde h  \overset{law}{=} \, \mathcal{N}(0, 8 |b+1/2|)  ,
\end{equation}
with $\mathcal{N}(\mu,\sigma^2)$ a Gaussian random variable with mean $\mu$ and variance $
\sigma^2$.  
\end{itemize}

\subsection{From the moments to the generating function as a Fredholm pfaffian}
\label{sec:Pfaffian} 

Here we will follow a route quite similar to \cite{GueudrePLD}, although more involved
because of the presence of boundary strings. We first use the Schuhr Pfaffian identity (see Appendix \ref{app:Fredholm} for the definition of
the Pfaffian and its properties)
\begin{align}
 & \prod_{1\leq i<j \leqslant n_s^0} D_{k_i^0,m_i^0,k_j^0,m_j^0}   \prod_{1\leq i<j \leqslant n_s} D_{k_i,m_i,k_j,m_j}  \prod_{i=1}^{n_s^0} \prod_{j=1}^{n_s} D_{k_i^0,m_i^0,k_j,m_j} \nn \\& =  \underset{2 (n_s+n_s^0) \times 2 (n_s+n_s^0)}{\rm Pf} \Big( \frac{X_i-X_j}{X_i+X_j} \Big)  \prod_{j=1}^{n_s } \frac{m_j}{2 i k_j}  \prod_{j=1}^{n^0_s } \frac{m^0_j}{2 i k_i^0} 
\end{align}
with $X_{2p-1} = m_p + 2i k_p$ and $X_{2p} = m_p - 2i k_p$ for $p\in [1,n_s ]$ and 
$X_{2n_s+2p-1} = m^0_p + 2i k_p^0$ and $X_{2n_s+2p} = m^0_p - 2i k_p^0$ 
for $p\in [1,n_s^0]$.  Using these identities, the definition of the generating function \eqref{defg0} and the formula for the moments $\overline{Z(t)^n}$ \eqref{EqT:SymmetrizedZn} we obtain the following expression for the generating function  
\begin{align}
& g_b(s) = \sum_{n_s\geq 0} \sum_{n_s^0 \geqslant 0} 
\frac{1}{n_s!}\frac{1}{n_s^0!}       \nn \\& \times 
\prod_{j=1}^{n_s^0}\sum_{(m_j^0,\ell_j^0)}^{(b)}  \int_\mathbb{R} \frac{\rmd k_j^0}{2\pi}   \frac{B_{k_j^0,m_j^0}}{4 i k_j^0} e^{ -  (t/4)^{\beta(b)} m_j^0 s}  (-1)^{m^0_j} b^{-2 m^0_j}  \nn \\& \times  (2i\pi) \left[ (k_j^0-k_{\ell_j,m_j^0}) \delta(k_j^0-k_{\ell_j,m_j^0}) -  (k_j^0+k_{\ell_j,m_j^0}) \delta(k_j^0+k_{\ell_j,m_j^0}) \right]\nn  \\& \times \prod_{j=1}^{n_s} \sum_{m_j \geqslant 1 }   \int_\mathbb{R} \frac{\rmd  k_j}{2 \pi} \frac{B_{k_j,m_j}}{4 i k_{j}}  e^{ -  (t/4)^{\beta(b)} m_j s}  (-1)^{m_j} b^{-2 m_j} 
\underset{2 (n_s+n_s^0) \times 2 (n_s+n_s^0)}{\rm Pf} \Big( \frac{X_i-X_j}{X_i+X_j} \Big)  \nn\\&
\times e^{-t \left[ \sum_{j=1}^{n_s^0}  E(m_j^0,k_j^0)  +  m_j  v_\infty(b) +  \sum_{j=1}^{n_s}  E(m_j,k_j)  + m^0_j v_\infty(b)\right] }
\end{align}
Now we use the following identity \be \label{identity} 
\frac{X_i-X_j}{X_i+X_j} = \int_{v_i , v_j >0}\rmd v_i \, \rmd v_j \,  2 \delta'(v_i-v_j) e^{-v_i X_i - v_j X_j},
\quad i,j\in [1,2 (n_s+n_s^0)].
\ee
and standard properties of the pfaffian to move the integrals outside the pfaffian
(see Appendix). This gives 
\begin{align}
& g_b(s) = \sum_{n_s\geq 0} \sum_{n_s^0 \geqslant 0} 
\frac{1}{n_s!}\frac{1}{n_s^0!}    \times \left(  
\prod_{j=1}^{n_s^0} \int_{v_{2j-1+2 n_s}>0, v_{2j +2 n_s}>0} \right)  \nn \\& \prod_{j=1}^{n_s^0}\sum_{(m_j^0,\ell_j^0)}^{(b)}  
\int_\mathbb{R} \frac{\rmd k_j^0}{2\pi}  \frac{ B_{k_j^0,m_j^0} }{4 i k_j^0} 
e^{ -  (t/4)^{\beta(b)} m_j^0 s}  (-1)^{m_j^0} b^{-2 m^0_j} 
   e^{- v_{2j-1+ 2n_s} (m^0_j + 2 i k^0_j) - v_{2j+ 2n_s} (m^0_j - 2 i k^0_j)}  
\nn \\&  {\blue (2i \pi) \left[ (k_j^0-k_{\ell_j,m_j^0}) \delta(k_j^0-k_{\ell_j,m_j^0}) -  (k_j^0+k_{\ell_j,m_j^0}) \delta(k_j^0+k_{\ell_j,m_j^0}) \right] } \nn \\
& \times \left(\prod_{j=1}^{n_s} \int_{v_{2j-1}>0, v_{2j}>0} \right)  \sum_{m_j \geqslant 1} \int_\mathbb{R} \frac{\rmd k_j}{2\pi}  \frac{ B_{k_{j},m_j} }{4 i k_j^0}  (-1)^{m_j} b^{-2 m_j} 
e^{  -  (t/4)^{\beta(b)} m_j s} e^{- v_{2j-1} (m_j + 2 i k_j) - v_{2j} (m_j - 2 i k_j)} \nn \\
& \times  
\underset{{2 (n_s+n_s^0) \times 2 (n_s+n_s^0)}}{\rm Pf}(2 \delta'(v_i-v_j) ) \nn\\&
\times e^{-t \left[ \sum_{j=1}^{n_s^0}  E(m_j^0,k_j^0)  + v_\infty(b) m_j+  \sum_{j=1}^{n_s}  E(m_j,k_j)  +  m^0_j v_\infty(b)\right] }
\end{align}
Note that this has the following structure:
\begin{align} \label{EqT:gbs01}
& g_b(s) = \sum_{n_s\geq 0} \sum_{n_s^0 \geqslant 0} 
\frac{1}{n_s!}\frac{1}{n_s^0!} \prod_{p=1}^{2 (n_s+n_s^0)}  \int_{v_p>0}\rmd v_p \, 
f_1(v_1,v_2) \cdots f_1(v_{2 n_s-1},v_{2 n_s})  \nn \\& 
\times f_2(v_{2 n_s+1},v_{2 n_s+2}) \cdots f_2(v_{2 n_s + 2 n_s^0 -1},v_{2 n_s+2 n_s^0}) 
  \times  
\underset{{2 (n_s+n_s^0) \times 2 (n_s+n_s^0)}}{\rm Pf}(\delta'(v_i-v_j) )
\end{align}
where we have introduced two antisymmetric functions $f_{1,2}(v_i,v_j)=- f_{1,2}(v_j,v_i)$ given by
\begin{equation} \label{deff1} 
f_1(v_i,v_j) = 
 \int_\mathbb{R} \frac{\rmd k}{2 \pi} \sum_{m \geqslant 1} 
 \frac{(-1)^{m}  b^{-2 m} B_{k,m}}{2 i k}
e^{-t  m k^2 + \frac{t}{12} m^3 -  (\frac{t}{4})^{\beta(b)} m s - m( v_i + v_j) - 2 i k (v_i-v_j)} e^{- (v_\infty(b)+\frac{1}{12})m t}
\end{equation}
where we recall that 
\be \label{B3} 
\frac{B_{k,m}}{2ik} =\frac{b^{2m}}{i\pi} \sinh(2 \pi k)  \Gamma(m+ 2 i k) \Gamma(m- 2 i k) 
 \frac{\Gamma\left(b + \frac{1-m}{2} - i k \right)  \Gamma\left(b + \frac{1-m}{2} + i k\right)}{ 
\Gamma\left(b + \frac{1+m}{2} - i k \right)  \Gamma\left(b + \frac{1+m}{2} + i k\right)}
\ee
and
\begin{align} \label{def2w1} 
& f_2(v_i,v_j) = 
\sum_{(m ,\ell)}^{(b)} (-1)^{m} 
\int_\mathbb{R} \frac{\rmd k}{2\pi} \frac{b^{-2 m} B_{k,m}}{2 i k}
e^{-t  m k^2 + \frac{t}{12} m^3 -  (t/4)^{\beta(b)} m s } e^{- m( v_i + v_j) - 2 i k (v_i-v_j)} \nn\\& \times  (2i \pi) \left[ (k-k_{\ell,m}) \delta(k-k_{\ell,m}) -  (k+k_{\ell,m}) \delta(k+k_{\ell,m}) \right] e^{-(v_\infty(b)+\frac{1}{12})m t}
 \, .
\end{align}
We can now "undo" our notational trick and obtain the following
definition of the antisymmetric function $f_2$ as
\bea 
\!\!\!\!\!\!\!\!\!\!\!\!\!\! \!\!\!\!\! && f_2(v_i,v_j) = 
\sum_{(m ,\ell)}^{(b)}  (-1)^m  \frac{i b^{-2 m}}{k_{\ell,m}} ({\rm res}_{k=k_{\ell,m}} B_{k,m}) e^{-t  m k_{\ell,m}^2 + \frac{t}{12} m^3 -  (\frac{t}{4})^{\beta(b)} m s } e^{-(v_\infty(b)+\frac{1}{12})m t} \nn \\
\!\!\!\!\!\!\!\!\!\!\!\!\!\! \!\!\!\!\!  && ~~~~~~~~~~~~~~~~~~~~~~~~~~~~~~\times e^{- m( v_i + v_j)}  \sin( 2 k_{\ell,m} (v_j-v_i) ) \label{deff2} 
\eea
where we recall that $k_{\ell,m} = i (b + \ell - \frac{m+1}{2})$ and (for integer $m$) 
\be \label{residue1} 
 \frac{i b^{-2 m}}{k_{\ell,m}} {\rm res}_{k=k_{\ell,m}} B_{k,m} =    \frac{i b^{-2 m} b_{k_{\ell,m},m}}{k_{\ell,m}} {\rm res}_{k=k_{\ell,m}} H_{k,m} = i   \frac{(-1)^{\ell-1} 2^{2m}  ( \frac{3}{2}-\ell-b)_{m-1} (1-\ell-b)_m}{(1-\ell-2 b)_{m} (\ell-1)! (m-\ell)!} \, ,
\ee
using \eqref{SH} and the second line in \eqref{bformula}.

\begin{remark} \label{r2} 
Note that exactly at the transition for the $m$-bound state, i.e. when $b+ \frac{1}{2} = \frac{m}{2}$
(see discussion below Eq. \eqref{45})
there is a removable singularity at $k=0$ in the $k$ integral for $f_1$ of the
type $\int_\mathbb{R} \rmd k \frac{\sin(2 k(v_i-v_j)}{k}$. Correspondingly there is no bound state exactly at
the transition. 
\end{remark} 

Using the symmetries of the integrand in \eqref{EqT:gbs01}
 it can be seen that we can finally rewrite $g_b(s)$ as
\begin{align} \label{EqT:gbs02}
& g_b(s)=   \sum_{N\geq 0}  \frac{1}{N!} 
\prod_{j=1}^{N}  \int_{v_j>0} \rmd v_j  
\underset{2N \times 2N }{\rm Pf}  \begin{pmatrix}
f_1(v_i,v_j) + f_2(v_i,v_j)  & 0 \\ 0 &  2 \delta'(v_i-v_j)  \end{pmatrix} \, .
\end{align}
Here we have used that for an arbitrary antisymmetric function $f(x,y)$ and integer $M$
\begin{align}
& \prod_{j=1}^{2 M} \int_{v_j>0} \rmd v_j \, f(v_1,v_2) \cdots f(v_{2M-1},v_{2M}) {\rm Pf}_{2 M \times 2M} (\delta'(v_i-v_j)) 
\nn \\&= \frac{2^M M!}{(2M)!} \prod_{j=1}^{2M} \int_{v_j>0}\rmd v_j \, \underset{2M \times 2M }{\rm Pf} (f(v_i,v_j))  {\rm Pf}_{2 M \times 2M} (\delta'(v_i-v_j)) \, ,
\end{align}
as can be easily checked using the definition of a Pfaffian,
$(2M-1)!!=\frac{(2M)!}{2^M M!}$ being the number of pairings of $2 M$ objects. 
Using this identity with $2M = N $ and $f=f_1+f_2$, the identity between \eqref{EqT:gbs01} and \eqref{EqT:gbs02} is easily checked term by term at $N=2 (n_s+n_s^0)$ fixed (notice that the terms with $N$ odd in \eqref{EqT:gbs02}  vanish). We also used that the product of two Pfaffians can be written as the Pfaffian of a block diagonal matrix. Finally, as is shown in the Appendix \ref{app:derpfaff} 
using properties of Fredholm Pfaffians, we obtain our main exact result for
the generating function, valid at arbitrary time $t$ and boundary parameter $b$
\begin{align} \label{mainresult1} 
g_b(s) = & \sqrt{ {\rm Det}[ I - P_0 {\cal K} P_0 ]  } \quad\quad\quad
{\cal K}(v_i,v_j) = \mathcal{K}_1(v_i , v_j) +  \mathcal{K}_2(v_i , v_j)
 \end{align}
 with 
 \begin{equation} \label{mainresult2}
\mathcal{K}_1(v_i,v_j) = 2 \partial_{v_i}   f_1(v_i,v_j),   \quad\quad  \mathcal{K}_2(v_i,v_j) = 2 \partial_{v_j}   f_2(v_i,v_j)  \, 
 \end{equation}
 and $P_y$ the projector on $[y,+\infty[$: $P_y(v_1,v_2) = \Theta(v_2-y) \delta(v_2-v_1)$. 
 We recall that the generating function was defined in \eqref{defg0}, and
 that the antisymmetric functions $f_1$ and $f_2$ are defined in 
 \eqref{deff1},\eqref{deff2}. Above and throughout this work ${\rm Det}[I+K]$ denotes the Fredholm determinant associated to a kernel 
 $K$ which is a linear operator acting on (a subset of) $\mathbb{L}^2(\mathbb{R})$. 
 
It is useful to note the following property. Suppose that the kernel 
${\cal K}$ is associated to $f$, i.e. ${\cal K}(v_i,v_j)= 2 \partial_{v_i}   f(v_i,v_j)$.
Denote $\tilde f(v_i,v_j)=f(a v_i,a v_j)$ with $a>0$. Then the kernel associated to 
$\tilde f$ is $\tilde{ {\cal K}}$ with $\tilde {\cal K}(v_i,v_j)= a {\cal K}(a v_i, a v_j)$.
Since it is obtained from ${\cal K}$ by a similarity transformation 
one has 
\be \label{simi} 
g_b(s) = \sqrt{{\rm Det}[ I - P_0 \tilde {\cal K} P_0 ]  }
\ee
Hence, the generating function is unchanged by any scale transformation on the
arguments of $f$. We will note below $\tilde {\cal K}$ the kernel with scaled arguments
(and respectively $\tilde {\cal K}_1$ and $\tilde {\cal K}_2$ their components).

\subsection{Matching with the result of Ref. \cite{AlexLD} for $b > - \frac{1}{2}$}
\label{matching} 

In the previous section we have obtained the generating function $g_b(s)$ in 
Eqs. \eqref{mainresult1}, \eqref{mainresult2} in terms of a kernel ${\cal K}$
involving the antisymmetric functions $f_1$ and $f_2$ defined in terms of formal sums over integers in 
\eqref{deff1},\eqref{deff2}. 
On the other hand in Ref. \cite{AlexLD} we have obtained 
\be \label{K0} 
g_b(s) = \sqrt{{\rm Det}[ I - P_0 K P_0 ]  } \quad , \quad K(v_i,v_j) = 2 \partial_{v_i} F(v_i,v_j) 
\ee 
in terms of an antisymmetric function $F$ expressed as
\begin{align}\label{eq:1DkernelAllT}
&F(v_i,v_j)=  \iint_{i \mathbb{R} + \kappa} 
\frac{\rmd w \rmd z}{(2i\pi)^2 }\frac{\Gamma(b+\frac{1}{2}-w)}{\Gamma(b+\frac{1}{2}+w)}\frac{\Gamma(b+\frac{1}{2}-z)}{\Gamma(b+\frac{1}{2}+z)}\\
&\hspace*{3cm}
\times 
\Gamma(2w)\Gamma(2z) \frac{\sin(\pi (z-w))}{\sin( \pi (w+z))}
e^{-v_i z-v_j w + t  \frac{w^3+z^3}{3}  - (w+z) s (t/4)^{\beta(b)} }  \nonumber
\end{align}
with $0< \kappa < \min(1,b + \frac{1}{2})$. This is obtained from (86-87) in Ref. \cite{AlexLD}
using the identity
\be
\int_\mathbb{R}\rmd r \frac{\varsigma}{\varsigma+ e^{-r}} \int \frac{\rmd w}{2 i \pi} f(w) e^{-w r}
= \int \frac{\rmd w}{2 i \pi} f(w) \varsigma^w \frac{\pi}{\sin \pi w} 
\ee 
with $\varsigma = e^{-s (t/4)^{\beta(b)}}$.

We now show that for $b \geq -1/2$ the two results match in the following sense.
If we rewrite $F$ as a sum over residues in $w$ and move the contour in $z$
appropriately we can rewrite $F$ as a formal sum over integer as
\be \label{id89} 
F(v_1,v_2) = f_1(\frac{v_1}{2},\frac{v_2}{2}) + f_2(\frac{v_1}{2},\frac{v_2}{2}) \quad, \quad b > -\frac{1}{2}
\ee

In order to show this let us close the contour on $w$ in \eqref{eq:1DkernelAllT} on the right, 
since $v_i,v_j>0$. The poles in $w$ are at $w = b + 1/2 + m$ (we denote them as poles of type I), 
and at $w+z=m$ (we denote as poles of type II). 

We can first show that the poles of type I have a vanishing contribution. Indeed the residues of these poles give $F$ as a sum with the following structure
\be
\sum_{m \geqslant 1} \int_{i \mathbb{R} + \kappa}  \frac{\rmd z}{2 i \pi} \frac{(-1)^{m+1}}{m!} \frac{\Gamma(1+ 2b + 2 m)}{\Gamma(1+ 2b + m)} 
\frac{\Gamma(2 z) \Gamma(b + \frac{1}{2} - z)}{\Gamma(b + \frac{1}{2} + z)} 
\frac{\sin \pi( z - b - \frac{1}{2} - m)}{\sin \pi( z + b + \frac{1}{2} + m)} \times \exp(...)
\ee 
Closing the contour on $z$ on the right, we see that 
poles of the Gamma function on the numerator at $z=b + 1/2 + m_2$ give a residue 
proportional to $\sin \pi(m_2-m)$ and therefore vanish. The poles of the 
the sine on the denominator are canceled by the $\Gamma(b + \frac{1}{2} + z)$. 

Thus we only need to consider poles of type II. For these, using $w=m-z$ we obtain
\bea
\fl && F(v_i,v_j)=\int_{i \mathbb{R} + \kappa} \frac{\rmd z}{2 i \pi} 
\sum_{\substack{m \geqslant 1 \\
m - {\rm Re}(z) >0}} 
 (-1)^{m+1}
\frac{\Gamma(b+\frac{1}{2} - m + z)}{\Gamma(b+\frac{1}{2} + m - z)}
\frac{\Gamma(b+\frac{1}{2} - z)}{\Gamma(b+\frac{1}{2} + z)}
 \nonumber \\
\fl && \times \Gamma(2 m - 2 z) \Gamma(2 z) e^{-  (t/4)^{\beta(b)}ms-v_i z - v_j (m-z) + \frac{t}{3} 
( z^3 + (m-z)^3)} \frac{\sin \pi(2 z-m) }{\pi}
\eea 
One then sets $i k = z - \frac{m}{2}$ with $i k \in i \mathbb{R} + \kappa - \frac{m}{2}$. 
\bea
\fl && F(v_i,v_j)= \int_{\mathbb{R} - i \kappa + i \frac{m}{2}} \frac{\rmd k}{2 \pi} 
\sum_{\substack{m \geqslant 1 \\
m/2 - {\rm Re}(i k) >0}} \frac{(-1)^{m+1}}{\pi}
\frac{\Gamma(b+\frac{1}{2} - m/2 + i k)}{\Gamma(b+\frac{1}{2} + m/2 - ik )}
\frac{\Gamma(b+\frac{1}{2} - i k - m/2)}{\Gamma(b+\frac{1}{2} + i k + m/2)}
 \nonumber \\
\fl && \times \Gamma( m - 2 i k ) \Gamma(2 i k + m) e^{-  (t/4)^{\beta(b)}ms-v_i(m/2+ i k ) - v_j (m/2-i k ) + \frac{t}{3} 
( (i k + m/2)^3 + (m/2-i k)^3)} \sin 2 \pi i k   \nonumber \\ \fl && 
\eea  
In order to show equivalence to the kernel $f_1 + f_2$ we now need to shift back $k$ to the real axis. To do that one notices that in the band $- \frac{m}{2} + \kappa \leqslant {\rm Re}(i k) \leqslant  0$ the only 
poles arise from the Gamma functions in the numerator and 
are located at $i k + b + 1/2 - m/2 = - n_1$ and $- i k + b + 1/2 - m/2 = - n_2$. Hence we must have
the following inequalities
\be \label{ineq} 
\frac{m}{2} - (b + \frac{1}{2}) \leqslant   n_1 < m - (b + \frac{1}{2}) \quad , \quad  - (b + \frac{1}{2}) < n_2  \leqslant 	 \frac{m}{2} - (b + \frac{1}{2}) 
\ee
One can check that the conditions for these poles to have non-zero residue (from the Gamma functions in the denominator) are $m>n_1,n_2$ which is automatically satisfied if \eqref{ineq} is obeyed.
\begin{remark} \label{r3} 
Note however that when $b+1/2 - m/2$ is a positive integer, the residue associated to $n_1=b+1/2 - m/2$
and the residue associated to $n_2=b+1/2 - m/2$ cancel one another. Hence we can make the inequality
strict. 
\end{remark} 
The shift of the contour thus gives
\bea
\fl && F(v_i,v_j)= F_1(v_i,v_j) + F_2(v_i,v_j) \\
\fl && F_1(v_i,v_j)= \int_{\mathbb{R} } \frac{dk}{2 \pi} 
\sum_{m \geqslant 1} \frac{(-1)^{m}  b^{-2 m} B_{k,m}}{2 i k}
e^{-  (t/4)^{\beta(b)}ms-v_i(m/2+ i k ) - v_j (m/2-i k )  -t  m k^2 + \frac{t}{12} m^3}  
\eea  
where $F_2$ is the sum over the residues, see below, and $B_{k,m}$ is defined 
in \eqref{B3} . We can already see, comparing 
with \eqref{deff1} and using that $v_\infty(b)=-\frac{1}{12}$ for $b \geqslant - \frac{1}{2}$, that 
\be
F_1(v_i,v_j)= f_1(\frac{v_i}{2},\frac{v_j}{2}) 
\ee
From the definition of $f_2$ in \eqref{def2w1},\eqref{deff2} as residues of $f_1$,
we see that in order to show that 
\be
F_2(v_i,v_j)= f_2(\frac{v_i}{2},\frac{v_j}{2}) 
\ee
one only needs to show that the allowed values of $m$, $n_1$ and $n_2$ correspond
exactly to the allowed values of $(m,\ell)$ which label the bound states.

We now explain the correspondence between these poles and the boundary bound states
which contribute to $f_2$. The index $m$ is the same in both cases, and we
set $\ell=n_2+1$. The residues at $n_1$ indicate the poles of $-k$ rather than $k$ so for simplicity we can reduce to the study of the poles at $n_2$. From the constraints
\eqref{ineq} plus Remark \ref{r3} (see also Remark \ref{r2}) the integer $\ell$ must satisfy
\be
\ell \geq 1 \quad , \quad m > 2 \ell + 2 b - 1
\ee 
This is clearly equivalent to the bound state structure presented in Table \ref{tab:stability1}
(see also \eqref{EqT:Existence}). This completes the identification \eqref{id89} for 
the unbound phase $b > -1/2$. Note that the factor of $2$ in the variables $v_i$ is
immaterial in the evaluation of the Fredholm determinant as can be seen using the similarity
transformation 
\eqref{simi} with $a=1/2$. Hence we have shown that the
main result \eqref{mainresult1}, \eqref{mainresult2} is formally
equivalent (at all time $t$) to the result of Ref. \cite{AlexLD} for $b > -1/2$.

\subsection{Discussion on the Mellin-Barnes summation formula for $b > - \frac{1}{2}$} 
\label{sec:MB} 

For other RBA solutions of the KPZ equation, in order to evaluate the formal sums over integers (arising from
the summation over the eigenstates of the LL Hamiltonian) one uses the so-called
Mellin-Barnes (MB) trick, which allows to rewrite the sum over the string lengths $m$ as an integration over a contour in the complex plane 
\begin{equation}\label{eq:MellinBarnesdef} 
\sum_{m \geqslant 1 } (-1)^m f(m) = - \int_{C} \frac{dz}{2 i \sin(\pi z)} f(z).
\end{equation}
where $C=a+  i \JR$ with $0<a<1$. However one fundamental assumption of the equality above is that the function $f(z)$ has no poles on the ${\rm Re}(z)>0$ half-plane. The kernel $f_1$ instead contains the function 
\be \label{EqT:defBkm-explicit}
B_{k,m} =\frac{2 kb^{2m} }{\pi} \sinh(2 \pi k)  \Gamma(m+ 2 i k) \Gamma(m- 2 i k) 
\frac{\Gamma\left(b + \frac{1-m}{2} - i k \right)  \Gamma\left(b + \frac{1-m}{2} + i k\right)}{ 
\Gamma\left(b + \frac{1+m}{2} - i k \right)  \Gamma\left(b + \frac{1+m}{2} + i k\right)}
\ee
It is easy to see that this functions has poles as function of $m$ at positions 
\begin{equation}
b + \frac{1-m}{2} \pm i k = - n
\end{equation}
with $n \geq 0$ a positive integer. Hence when performing MB one must take into account
these extra poles. As we have shown in the previous section, these extra poles $n>0$ correspond exactly to the bound state contributions carried by the kernel $f_2$. Backtracking
the steps performed in the previous section allows to perform the correct MB summation
for $b > -1/2$. Performing the MB summation for $b< -1/2$ is left for a future study. 

%
%
%
%
%

\subsection{Large time limit: unbound phase, $b > -1/2$} 
\label{sec:res1} 

In a previous section we have obtained the generating function $g_b(s)$ at all times in 
Eqs. \eqref{mainresult1}, \eqref{mainresult2} in terms of a kernel ${\cal K}$.
Then we have shown that for $b > -1/2$ this kernel is identical, up to a similarity transformation
that leaves the Fredholm determinant unchanged, to the kernel $K$ of Ref. \cite{AlexLD},
${\cal K}(v_i,v_j) = \frac{1}{2} K(\frac{v_i}{2},\frac{v_j}{2})$.
As it was shown in \cite{AlexLD} the generating function obtained from the 
kernel $K$ defined in \eqref{K0}, \eqref{eq:1DkernelAllT}, converges 
in the large time limit to the TW-GSE distribution for $b>-1/2$ and to the TW-GOE distribution at $b=-1/2$.
Setting
\begin{equation}
b +1/2 = \epsilon t^{-1/3}
\end{equation}
and taking the large time limit at fixed $\epsilon$,
allows to explore the critical crossover region by varying $\epsilon$ from $0$ to $+\infty$. 
The transition kernel is obtained by setting $\beta(b) =1/3$ for $b \geqslant - 1/2$ and rescaling in \eqref{eq:1DkernelAllT} the variables as 
\begin{align}
w \to {w} t^{-1/3} \\
z \to {z} t^{-1/3} \\
(v_i,v_j) \to ({v}_i t^{1/3},{v}_j t^{1/3})  
\end{align}
Taking the large time limit of the kernel one obtains indeed 
\begin{align} 
&\lim_{t \to \infty} K({v}_i t^{1/3}, {v}_j  t^{1/3}) = 2 \lim_{t \to \infty} t^{-1/3}\partial_{{v}_i} F({v}_i t^{1/3},{v}_j t^{1/3}) \\
& \lim_{t \to \infty} F({v}_i t^{1/3},{v}_j t^{1/3})=  \iint_{i \mathbb{R} + A_\epsilon}
\frac{\rmd w \rmd z}{(2i\pi)^2 }\frac{\epsilon+w  }{ \epsilon-w}\frac{\epsilon+z }{  \epsilon-z}
\frac{1}{4wz} \frac{z-w}{w+z}
e^{- (w+z) s /2^{2/3}-v_i z- v_j w +  \frac{w^3+z^3}{3} } \nonumber
\end{align}
where $A_\epsilon \in [0,\epsilon]$. 
\begin{itemize}
\item In the limit $\epsilon \to + \infty$ one recovers the standard GSE kernel, therefore giving 
\be \label{mainresult12} 
\lim_{\epsilon \to + \infty} \lim_{t \to \infty} g_b(s)=  \lim_{\epsilon \to + \infty} \lim_{t \to \infty} \sqrt{ {\rm Det}[ I- P_0 {\cal K} P_0 ]  } 
=\sqrt{ {\rm Det}[ I - P_{2^{-2/3 }s} { K}^{\rm GLD} P_{2^{-2/3} s} ]  } 
\ee
with ${ K}^{\rm GLD} (v,w) = K_{\rm Ai}(v,w) - \frac{1}{2} \Ai(v) \int_0^\infty dy \Ai(y + w)$.
\item The other limit, $\epsilon \to 0^+$, requires particular care. First the integration contour for $z$ and $w$ need to be shifted to the right of $\epsilon + i \mathbb{R}$. Then the limit $\epsilon \to 0^+$ can be taken, as shown in  \cite{AlexLD}, giving the scalar form of the GOE kernel 
\begin{equation}
\lim_{\epsilon \to + 0^+} \lim_{t \to \infty} g_b(s)=\sqrt{ {\rm Det}[ I - P_{2^{-2/3 }s} { K}^{\rm GOE} P_{2^{-2/3} s} ]  } 
\end{equation}
with ${ K}^{\rm GOE} (v,w) = K_{\rm Ai}(v,w) + \Ai(v ) - \Ai(v) \int_0^\infty dy \Ai(y + w)$.
\end{itemize}

 \subsection{Large time limit: bound phase $b<-1/2$}\label{sec:attractivebounfdaryGOE}
 
Let us discuss now the bound phase, $b<-1/2$. The formula Eqs. \eqref{mainresult1}, \eqref{mainresult2}
for the generating function $g_b(s)$ at all times in terms of the kernel ${\cal K}$ remains correct. 
Unfortunately, the summation over all eigenstates (i.e. all integers $m,\ell$) remains at present
our of reach. However, inside the bound phase, it reasonable to expect that the large time
limit is given by the contribution of the bound state to the wall formed of a single boundary string with $\ell=1$. As we show below it predicts a
Gaussian distribution for the fluctuations of the free energy of the directed polymer (i.e. of
the KPZ height) at the origin at large time. Such a distribution is expected from universality and the results 
on the discrete models, e.g. \cite{BaikSymPermutations}.

A physical argument can be given. For $b<-1/2$ this state is the ground state of the LL model for any $n$. The reason why restricting to the ground state does not reproduced the full physics in the case $b \geqslant -1/2$ is because the limit $L \to \infty$ is taken before the $t \to \infty$ limit. In the quantum language the phase $b \geqslant  -1/2$ is gapless and the full spectrum of excitations contributes to the fluctuations of $\tilde{h}$ (for examples pure bulk string states with $n_s=1$ and $n_s^0=0$ and a small string momenta $k \sim 1/L$ have an energy that is arbitrarily close to the ground state). On the other hand when $b <-1/2$ the system is gapped. The effect of taking $L \to \infty$ does not change the problem, since the polymer remains bound to the wall and its typical wandering away from it is $\mathcal{O}(1)$. By contrast, in the $b \geqslant -1/2$ phase it explores distances to the wall of order $L^{2/3}$.\\

We thus start back from the initial expression of the generating function $g_b(s)$ \eqref{defg0}, 
with $\beta=1/2$, $v_\infty(b)= (b + \frac{1}{2})^2 - \frac{1}{12}$ from \eqref{vinfty}, and
using \eqref{EqT:SymmetrizedZn} 
for $\overline{Z(t)^n}$ and retaining only the contribution from the ground state
, which, as was discussed in section \ref{sec:GSLLb} is a boundary string $n_s=0$, $n_s^0=1$, $m^0=n$ and $\ell=1$. We obtain using \eqref{residue1} 
\bea
\fl && g_{b<-1/2}(s) =  1+ \sum_{n\geqslant 1} \frac{(-1)^n}{n!} e^{-\frac{1}{2} \sqrt{t} s n}\overline{Z(t)^n e^{  t v_\infty(b)n }}
\nn \\
\fl && 
=\sum_{n\geq 0}  \frac{ {(-1)^n}{}  2^{2n-1} }{n \Gamma(n)} e^{-\sqrt{t}/2 s n} e^{-t (  ( b+\frac{1}{2}) n^2 -\frac{n^3}{3})} (n - 1 - 2 b) \frac{\Gamma(n - b)}{\Gamma(-b)} \frac{ \Gamma( - 2 b)}{  \Gamma(n-2 b)}  \frac{\Gamma(n - b - \frac{1}{2})}{\Gamma(\frac{1}{2}-b)} \nn
\eea
 We  use the Gaussian decoupling 
\begin{equation}
e^{t | b + \frac{1}{2}| n^2 } =  \pi^{-1/2} \int_{\mathbb{R}} \rmd y e^{-  {y^2}{} + n \sqrt{4 t |b+ \frac{1}{2}|} y} ,
\end{equation}
and a MB formula similar to \eqref{eq:MellinBarnesdef} for the sum over $n \geqslant 0$
\begin{equation}\label{eq:MellinBarnesdef2} 
\sum_{n \geqslant 0 } (-1)^n f(n) = - \int_{C} \frac{dz}{2 i \sin(\pi z)} f(z).
\end{equation}
where $C=a+  i \JR$ with $-1<a<0$. To obtain the large time limit, 
the proper rescaling for the variable $z$ is not given by $1/\lambda=(4/t)^{1/3}$ but by $1/\sqrt{t}$. 
We thus write $z \to z/\sqrt{t}$, in which case the cubic term disappears at large time, leading to
\begin{equation}
 \lim_{t \to \infty}  g_b(s) =    - \sqrt{\frac{1}{\pi}} \int_\mathbb{R} \rmd y \int_{0^- + i \JR} \frac{\rmd z}{ 2 \pi i  z}  e^{-  {y^2}{} + z \sqrt{4  |b+ \frac{1}{2}|} y}      e^{- s z/2}     \nn
\end{equation}
\begin{equation}
=     \sqrt{\frac{1}{\pi}} \int_\mathbb{R} \rmd y    e^{-  {y^2}{}} \Theta(s/2   - \sqrt{4  |b+ \frac{1}{2}|} y ) =       \sqrt{\frac{1}{2 \pi}} \int_{-\infty}^{\frac{s   }{\sqrt{8 |b + \frac{1}{2}|}}} dy    e^{-  \frac{y^2}{2}} .
\end{equation}
We thus obtain a Gaussian distribution for the rescaled height $\tilde{h}$
\begin{equation}
\lim_{t \to \infty}  \mathbb{P}(\tilde h < s) = \lim_{t \to \infty}  g_b(s) = 
{\frac{1}{\sqrt{2\pi 8 |b+1/2|} }} \int_{-\infty}^s dy    e^{-  \frac{y^2}{16 |b+1/2|}}  \, ,
\end{equation}
showing that $\tilde{h}$ is a Gaussian random variable of variance $8|b+1/2|$.

\section{Conclusions and future directions}
In this work we have conducted a systematic analysis of the attractive Lieb-Liniger model on the half-line. In the closely related XXZ spin chain problem it corresponds to open diagonal boundary conditions.
In particular we have characterized its spectrum of boundary bound states as a function of the particle number and of the interaction with the boundary. We expect that these results may be of interest in the context of quantum quenches in restricted geometries and non-linear Schrodinger systems. \\

These results have been used to obtain predictions via the replica Bethe ansatz for the fluctuations of the KPZ equation next to a wall. The height of the KPZ function, which also maps to the free energy of the directed polymer in presence of disorder on the half space, displays a phase transition at the critical value of the interaction with the wall. We have obtained the three types of statistics that govern the fluctuations in the different regimes: attractive boundary, critical boundary and unbounded growth, described respectively by the Gaussian, GOE-TW and GSE-TW distributions.
Moreover we quantify the KPZ front velocity. The present results complement the
ones obtained in Ref. \cite{AlexLD} {and are in agreement with results previously obtained in
a number of works on discrete models, thereby consistent with the universality of the transition.}
\\
Although the exact formula for the integer moments of the partition sum of the directed polymer
have been obtained here for all $b$ and time $t$, extracting the PDF (i.e. the generating function at all time) has been possible only in the unbound phase {$b > - 1/2$} and at the transition.
One of the remaining open questions is to obtain a complete solution for the bound phase at
arbitrary time, e.g. by developing a specific Mellin-Barnes method. Obtaining the
GOE-Gaussian crossover kernel in the large time limit (for $\epsilon<0$) for the
KPZ equation would be of great interest and could be compared with
the results of Ref. \cite{BarraquandSchur} for the discrete model.
By duality, the case $b<-1/2$ amounts to study a Brownian IC with a
positive drift, which is also of interest: this would result in a better understanding of the summation method which could then probably 
be extended to the full space case. In addition, results for endpoints 
away from the wall would also be desirable.

%
 
Finally, the same type of phase transition is expected to take place in the quantum regime, in particular in cold atomic systems which are governed by the attractive Lieb-Liniger Hamiltonian. It will be fascinating to experimentally observe a similar localization transition around a point-like defect, for example induced by a local laser field, simply by tuning its intensity.

\section*{Acknowledgments}

We are grateful to Guillaume Barraquand for very useful remarks and pointing out references. We also
thank A. Bufetov, I. Corwin, A. Borodin, K. Takeuchi, N. Zygouras for discussions.
This work is supported in part by LabEX ENS-ICFP:ANR-10-LABX-0010/ANR-10-IDEX-0001-02 PSL* and Research Foundation Flanders (JDN and TT), ANR grant ANR-17-CE30-0027-01 RaMaTraF (PLD and AK),
{ERC under Consolidator grant (AK)} number 771536 (NEMO)  
and the National Science Foundation 
under Grant No. NSF PHY11-25915 
during the randomKPZ16 program at the KITP.

\newpage

\begin{appendices}

\section{Number of boundary strings $N(b,m^0)$}
\label{app:number} 

The number of boundary strings $N(b,m^0)$ for a fixed value of $m^0 \geqslant 1$ is equal to
the number of values of $\ell$ possible for a given $m^0$. From the Table \ref{tab:stability1} we find, denoting ${\rm E}(x)$ the integer part of $x \geqslant 0$

\begin{itemize}

\item 
For $b \geqslant 0$, we have, for ${\rm E}(\frac{2b}{\bar c}) \leqslant m^0-1$
\be
N(b,m^0) = {\rm E}(\frac{m^0 - {\rm E}(2 \frac{b}{\bar c})}{2}) = 
\begin{cases} 
\frac{m^0}{2} - {\rm E}(\frac{b}{\bar c} + \frac{1}{2}) \quad \text{for} \quad m^0 \quad \text{even}  \\ 
\frac{m^0-1}{2} - {\rm E}(\frac{b}{\bar c})  \quad ~~~ \text{for}  \quad m^0 \quad \text{odd}
\end{cases}
\ee
and $N(b,m^0) =0$ for ${\rm E}(\frac{2b}{\bar c}) \geqslant m^0$.

\item
For $b<0$, and $-2 b$ not an integer\footnote{From Table~\ref{tab:stability1}, we have $p={\rm E}(- \frac{2b}{\bar c})$ 
and for $p \geqslant 1$, 
$N(b,m^0)$ is one plus the number of integers in the interval $[1+p,\min(m^0,\frac{m^0+p+1}{2})]$}
, we have $N(b,m^0)={\rm E}(\frac{m^0+1}{2})$ for ${\rm E}(- \frac{2b}{\bar c})=0$, then
for $1 \leqslant {\rm E}(- \frac{2b}{\bar c}) \leqslant m^0-1$
\be
N(b,m^0) = \begin{cases} 
1 + \frac{m^0 - {\rm E}(- \frac{2b}{\bar c})}{2} \quad , \quad m^0 - {\rm E}(- \frac{2b}{\bar c}) \quad \text{even} \\
\frac{3}{2} + \frac{m^0 - {\rm E}(- \frac{2b}{\bar c})}{2} 
 \quad , \quad m^0 - {\rm E}(- \frac{2b}{\bar c}) \quad \text{odd}
 \end{cases}
\ee
with $N(b,m^0)=2$ for ${\rm E}(- \frac{2b}{\bar c})=m^0-1$, 
$N(b,m^0)=1$ for ${\rm E}(- \frac{2b}{\bar c})=m^0$, and finally $N(b,m^0)=0$ for 
${\rm E}(- \frac{2b}{\bar c}) \geqslant m^0+1$. 

\end{itemize}

\section{Lieb-Liniger model on the half line as the scaling limit of an open XXZ spin-$1/2$ chain with longitudinal fields at the boundaries} \label{app:XXZ}
We consider the XXZ spin 1/2 chain with open boundary conditions. In particular we focus on the case where at the two boundaries two longitudinal fields are applied, so-called diagonal boundary conditions. The Hamiltonian is given by 
\begin{equation}\label{eq:XXZopen1}
H = \sum_{j=1}^{L-1} \left( \sigma_j^x \sigma_{j+1}^x + \sigma_j^y \sigma_{j+1}^y + \Delta \sigma_j^z \sigma_{j+1}^z \right) + h_- \sigma_1^z + h_+ \sigma_L^z,
\end{equation}
with $\Delta = \cos \eta$ and with the longitudinal fields parametrised as $h_\pm = \sinh \eta \coth \xi_\pm$. Its diagonalization can be performed by a modified version of the
algebraic Bethe Ansatz proposed by Sklyanin in \cite{Sklyanin} and some properties of its correlation functions are known \cite{mailletterras1,mailletterras2}. The resulting Bethe equations reads as 
\begin{align}\label{eq:BetheXXZOpen}
\left( \frac{\sinh(\lambda + \eta/2)}{\sinh (\lambda - \eta/2)} \right)^{2L}&  \frac{\sinh(\lambda_j + \xi_+ - \eta/2)\sinh(\lambda_j + \xi_-  - \eta/2)}{\sinh(\lambda_j - \xi_+ +\eta/2)\sinh(\lambda_j - \xi_-  + \eta/2)} \nn \\&=  \prod_{k\neq j} \frac{\sinh(\lambda_j - \lambda_k + \eta) \sinh(\lambda_j + \lambda_k + \eta)}{\sinh(\lambda_j - \lambda_k - \eta) \sinh(\lambda_j + \lambda_k - \eta)}.
\end{align} 
The Lieb-Liniger model on the half line is given by the model \eqref{eq:XXZopen1} with the following scaling limit $\epsilon \to 0$ \cite{scalingXXZLL} 
\begin{align}
&\eta = i \pi - i \epsilon, \\
& L =  L^{LL} c/\epsilon,\\
& \lambda = \lambda^{ LL} \epsilon/{c},\\
& \xi_\pm = i \pi/2  - i \epsilon/2 ( 1 - 2 b_\pm/{{c}}).
\end{align}
In this limit the Bethe equations \eqref{eq:BetheXXZOpen} become the Bethe equations for the Lieb-Liniger model on the half-line with boundaries parametrized by $b_-$ and $b_+$ and with attractive coupling $\bar{c} = - c$ \eqref{EqT:BetheEq1}
\be 
e^{2 i \lambda^{LL}_\alpha L^{LL}} = 
\frac{b_+ - i \lambda^{LL}_\alpha}{b_++ i \lambda^{LL}_\alpha}  \frac{b_- - i \lambda^{LL}_\alpha}{b_- + i \lambda^{LL}_\alpha} \, \prod_{1 \leqslant \beta \neq \alpha \leqslant n} \frac{\lambda^{LL}_\alpha - \lambda^{LL}_\beta - i \bar c}{\lambda^{LL}_\alpha - \lambda^{LL}_\beta + i \bar c} 
\frac{\lambda^{LL}_\alpha + \lambda^{LL}_\beta  - i \bar c}{\lambda^{LL}_\alpha  + \lambda^{LL}_\beta +  i \bar c}.  \nn 
\ee

\section{Norms of the string states} \label{app:normStrings}

We detail in this appendix the derivation of the norm of the string states \eqref{EqT:NormFinal} starting from the finite size norm formula \eqref{EqT:normGaudin}. The non-trivial aspect of this calculation amounts in principle in evaluating the determinant in \eqref{EqT:normGaudin}. The symplifying feature that allows an exact formula is that both the prefactor and the determinantal part in \eqref{EqT:normGaudin} are singular for string states with rapidites of the form \eqref{EqT:BoundaryStringRapidities} (boundary strings) or \eqref{EqT:UsualString} (bulk strings): the prefactor tends to zero with the string deviations while the determinant diverges. This means that in the large $L$ limit it is sufficient to evaluate the most singular term of the determinant which, combined with the trivial prefactor, gives the exact result. For a state of $n$ particles made of $n_s$ bulk strings and $n_s^0$ boudnary strings, it can be seen that this most singular term factorizes among the different strings (divergent terms inside the determinant only appear within blocks inside a string). For this reason we evaluate here the norm for the special cases of a state made of (i) a single bulk string; (ii) a single boundary string. The final formula \eqref{EqT:NormFinal} then easily follows using the factorization property.

\subsection{Norm of a propagating string}

The above eigenfunctions are not normalized. We conjecture, that their norm is a generalization of the full space Gaudin determinant \cite{gaudin2014bethe}, given by
\bea \label{EqT:normGaudin}
\!\!\!\!\!\!\!\!\!\!\!\!\!\!\!\!\!\!\!\!\!\!\!\!\!\!\!\! || \Psi_\mu||^2= && \frac{n!}{4^n} \prod_{1\leq \alpha < \beta \leqslant n} \left( 1 + \frac{ \bar c^2}{(\lambda_\alpha - \lambda_\beta)^2} \right) \prod_{1\leq \alpha < \beta \leqslant n} \left( 1 + \frac{\bar c^2}{(\lambda_\alpha + \lambda_\beta)^2} \right) \prod_{\alpha=1}^n \left( 1 + \frac{\lambda_\alpha^2}{b^2} \right) \nn \\ 
&& \! \! \! \! \! \! \! \! \! \! \!  \! \! \!  \times \det_{n \times n} \left( \delta_{\alpha \beta} \Big( 2L + K_b(\lambda_\alpha)  
+  \sum_{1 \leqslant \gamma \neq \alpha \leqslant n} K_{+}(\lambda_\alpha , \lambda_\gamma) \Big) 
- (1-\delta_{\alpha \beta}) K_{-}(\lambda_\alpha , \lambda_\beta) \right)
\eea
where 
\begin{equation}
K_{\pm}(\lambda, \mu) =  -\left(\frac{2 \bar c }{\bar c^2+(\lambda -\mu)^2} \pm
\frac{2 \bar c}{\bar c^2+(\lambda +\mu )^2} \right) 
\end{equation} 
and
\begin{equation}
K_b(\lambda) = \frac{2 b}{b^2+\lambda ^2} \, .
\end{equation}

We comment on the rationale behind this conjecture in ~Appendix \ref{app:normGaudin}.
We note that this formula agrees with the conjecture already presented in \cite{Castillo} and on the scaling limit (see Appendix \ref{app:XXZ}) of the norm of the open XXZ eigenstates \cite{mailletterras1} (formula 4.26 in that paper).

We start here with the case of a single bulk string with rapidities $\lambda_a = k +  \frac{ i \bar c}{2} (m+1 - 2 a)$ in the $L=\infty$ with $a = 1 \dots , m$ and $m=n$. We reintroduce explicitly the string deviations $\delta$
as follows: $\lambda_a = k + \frac{ i \bar c}{2} (m+1 - 2 a) + i \delta_a$ and note $\delta_{a,a+1} = \delta_a- \delta_{a+1}$. We have introduced a $i$ factor in front of the string deviations compared to \eqref{EqT:UsualString} to make the parallel with \cite{cc-07} transparent.
Two terms appearing in the prefactor of the norm formula \eqref{EqT:normGaudin} are non singular:
\bea
&&  \prod_{i<j=1}^m \left( 1 + \frac{\bar c^2}{(\lambda_ i + \lambda_j)^2} \right) =  \frac{(2 k+i \bar{c} m) \left(\frac{i k}{\bar{c}}-\frac{m}{2}+\frac{1}{2}\right)_m}{2 k \left(\frac{i k}{\bar{c}}-\frac{m}{2}\right)_m} \nn \\
 &&  \prod_{j=1}^m  \left( 1 + \frac{\lambda_j^2}{b^2} \right) = \left(-\frac{\bar{c}^2}{b^2}\right)^m \left(\frac{-2 b+\bar{c}+2 i k-\bar{c} m}{2 \bar{c}}\right)_m \left(\frac{2 b+\bar{c}+2 i k-\bar{c} m}{2 \bar{c}}\right)_m \nn
\eea
Note that in the text we use the notation $H_{k,m}= \prod_{j=1}^m  ( 1 + \frac{\lambda_j^2}{b^2} )^{-1}$.
The singular term is evaluated, to leading order in the string deviation, as 
\bea
\prod_{i<j=1}^m \left( 1 + \frac{\bar{c}^2}{(\lambda_ i - \lambda_j)^2} \right)  \simeq \frac{m}{\bar{c}^{m-1}} \prod_{a=1}^{m-1} \delta_{a,a+1}
\eea
This is cancelled by a divergence coming from the Gaudin determinant:
\bea
\det (G) \simeq \frac{1}{\prod_{a=1}^{m-1} \delta_{a,a+1}} 2 m L + \mathcal{O}(1) \, .
\eea
This follows by proceeding exactly as in Appendix B of \cite{cc-07} to which we refer for more details and here we recall the main idea: divergent terms inside the determinant are on the three main diagonals of the Gaudin matrix and are those of the form $K_{\pm}(\lambda_{a},\lambda_{a+1})  \sim \frac{1}{\delta_{a,a+1}}$ with $a=1,\cdots , m-1$. . All these terms can be put onto the diagonal by applying the following transformation on the matrix: first add the first column to the second one then the first row to the second row (tranformations that leave the determinant unchanged). Then the term $\sim \frac{1}{\delta_{1,2}}$  only appears inside the $(1,1)$ entry of the matrix. Now add the second column to the third one and the second row to the third one. Then the term $\sim \frac{1}{\delta_{2,3}}$  only appears inside the $(2,2)$ entry of the matrix. Proceed recursively. Once all terms are on the diagonal, the determinant is evaluated by taking them. The $(m,m)$ entry of the matrix is also automatically taken and is to leading order equal to $2m L$ in the large $L$ limit. This leads to the above result. Taking all terms into account, the norm of the string state is evaluated as
\bea
|| \Psi_\mu ||^2 = \frac{L m!}{4^m} \frac{2 m^2 }{\bar{c}^{m-1}}    \frac{(2 k+i \bar{c} m) \left(\frac{i k}{\bar{c}}-\frac{m}{2}+\frac{1}{2}\right)_m }{ 2 k \left(\frac{i k}{\bar{c}}-\frac{m}{2}\right)_m }   \nn \\
\times \left(-\frac{\bar{c}^2}{b^2}\right)^m \left(\frac{-2 b+\bar{c}+2 i k-\bar{c} m}{2 \bar{c}}\right)_m \left(\frac{2 b+\bar{c}+2 i k-\bar{c} m}{2 \bar{c}}\right)_m  \, ,
\eea
which can be seen to be in agreement with \eqref{EqT:NormFinal}, i.e. for such a single bulk string state
with $m=n$ and $n_s=1$, one has
$1/|| \Psi_\mu ||^2 = \frac{4^m}{L m!} S_{k,m} H_{k,m}$ (there are no inter-string factors $D$, see below).

\subsection{General case}

The general case of a state made of $n_s$ bulk strings and $n_s^0$ boundary strings easily follows from the factorization property of diverging terms inside the determinant. The careful evaluation of the prefactor of \eqref{EqT:normGaudin} for such a state leads to the inter-string factors noted as $D$ in \eqref{EqT:defDSK}.
{ Note that for boundary strings, we have noticed that the conjectured formula misses a factor $(-1)^{m^0}$ per each boundary string, which we have corrected
for the formula we use, eq. \eqref{EqT:NormFinal}.}

\section{Gaudin determinant for the norm of Bethe states on the half-line } \label{app:normGaudin}

The logic behind the conjecture \eqref{EqT:normGaudin} follows from the study of other models that are integrable by coordinate Bethe ansatz: inside the determinant is inserted the matrix that is the derivative of the logarithm of the Bethe equations \cite{KorepinBook}: 
\bea
&& 
 \delta_{\alpha \beta} \Big( 2L + K_b(\lambda_\alpha)  
+  \sum_{1 \leqslant \gamma \neq \alpha \leqslant n} K_{+}(\lambda_\alpha , \lambda_\gamma) \Big) 
- (1-\delta_{\alpha \beta}) K_{-}(\lambda_\alpha , \lambda_\beta) = \nn \\
&& \frac{1}{i}  \partial_{\lambda_\beta} \log \left( e^{2 i \lambda_\alpha L} \left[ \prod_{\gamma \neq \alpha} \frac{\lambda_\alpha - \lambda_\gamma - i \bar c}{\lambda_\alpha - \lambda_\gamma  + i \bar c} 
\frac{\lambda_\alpha + \lambda_\gamma   - i \bar c}{\lambda_\alpha  + \lambda_\gamma  +  i \bar c} \frac{b - i \lambda_\alpha}{b+ i \lambda_\alpha}  \right]^{-1} \right)  \, . 
\eea
Notice that this formula was already conjectured in \cite{Castillo}.  
The prefactor in front of the determinant is then guessed by expanding formally the norm as a double sum over permutations for real rapidities (corresponding to a state with $n_s=n$, i.e. 
with $n$ one-strings (i.e. $n$ "particles") in the large $L$ limit) as, schematically, taking the modulus square of \eqref{EqT:wave} 
\bea
\!\!\!\! \!\! \!\! ||\Psi_{\mu} ||^2 = \frac{1}{4^n} \int_{0 \leqslant x_1 \leqslant \cdots \leqslant x_N \leqslant L}\sum_{P} \sum_{\epsilon} \sum_{P'} \sum_{\epsilon'} (\prod \epsilon_i) A  (\prod \epsilon'_i) (A')^* e^{i(\dots)} e^{-i(\dots)} \, , \nn
\eea
and to leading order in $L$, the only terms that survive are those with $P=P'$ and $\epsilon_i = \epsilon'_i$. This leads to
\bea 
\!\!\!\!\!\!\!\!\!\!\!\!\!\!\!\!\!\!\!\!\!\!\!\!\!\!\!\!\!\!\!\!\!\!\!\!\!\!\! ||\Psi_\mu ||^2 = \frac{ n!}{4^n} 2^n L^n\prod_{1\leq \alpha < \beta \leqslant n} \left( 1 + \frac{ \bar c^2}{(\lambda_\alpha - \lambda_\beta)^2} \right) \prod_{1\leq \alpha < \beta \leqslant n} \left( 1 + \frac{\bar c^2}{(\lambda_\alpha + \lambda_\beta)^2} \right) \prod_{\alpha=1}^N  \left( 1 + \frac{\lambda_\alpha^2}{b^2} \right)  , \nn \\
\eea
with corrections of order $\mathcal{O}(L^{n-1})$. 
This expression is indeed the leading order in $L$ obtained from the computation of the Gaudin formula \eqref{EqT:normGaudin} with the prefactor as above.

\section{Pfaffians, Fredholm Pfaffians and Fredholm determinants}\label{app:Fredholm}

\subsection*{Pfaffians} 

Let us first recall the definition of the Pfaffian of an antisymmetric matrix $A$ of size
$2n \times 2n$
\be
{\rm Pf} A = \sum_{\substack{\sigma \in S_{2n},\\ \sigma(2j-1) <\sigma(2 j)}} (-1)^\sigma \prod_{i=1}^n A_{\sigma(2i-1),\sigma(2i)} 
\ee
with $({\rm Pf} A)^2 = \det A$. Since the determinant of an antisymmetric matrix of odd size is zero,
the Pfaffian can also be set to zero. For $n=1$ it is ${\rm Pf} A= A_{12}$, for $n=2$ it is
${\rm Pf} A= A_{12} A_{34}-A_{13} A_{24}+ A_{14} A_{23}$, and so on, the 
number of terms in the sum is $(2n-1)!!$, the number of pairings of $2n$ objects. 
In the text we use the Schur Pfaffian identity
\be
\underset{2n \times 2n}{{\rm Pf}}\left(\frac{X_i-X_j}{X_i+X_j}\right) = \prod_{1 \leqslant i < j \leqslant 2n} \frac{X_i-X_j}{X_i+X_j}
\ee 

\subsection{Fredholm Pfaffians and Fredholm determinants} 

We here shortly review how Fredholm determinants and Pfaffians are defined. 
Given a kernel $K(u,v)$ and a domain $X \in \mathbb{R} $ we define a trace-class operator $K(u,v)$ satisfying   
\begin{equation}
\text{Tr}[K]= \int_X \rmd z K(z,z) <  \infty .
\end{equation}
Its Fredholm determinant is defined as the infinite convergent sum
\begin{equation}\label{eqn:FD1}
\text{Det}( I+  P_X K  P_X) = \sum_{n=0}^\infty \frac{1}{n!} \left( \prod_{j=1}^n \int_X \rmd z_j   \right) \det_{i,j=1}^n K(z_i,z_j) ,
\end{equation}
where we introduced the operator $P_X$ as a projector on the domain $X$ and the identity operator $I$. Following Ref.~\cite{Bornemann} we can see the Fredholm determinant as a discretization of the domain $X$ in $m$ points $\{ x_j \}_{j=1}^m$ with $m$ associated weights $\{ w_j \}_{j=1}^m$ such that given any integrable function $f(x)$ on $X$ we have
\begin{equation}
\int_X \rmd y K(x_i,y) f(y) = \lim_{m \to \infty} \sum_{j=1}^m K(x_i, x_j) f(x_j) w_j ,
\end{equation}
then the Fredholm determinant \eqref{eqn:FD1} is the limit of the determinant of the $m\times m$ matrix $\delta_{ij}+ [K](x_i,x_j) $
\begin{equation}
\text{Det}( I+  P_X K  P_X)  = \lim_{m \to \infty} \det \left( \delta_{ij} +[K](x_i,x_j)  \right)_{i,j=1}^m .
\end{equation}

A generalization of the Fredholm determinant is given by the Fredholm Pfaffian defined for antisymmetric matrices. Given a two-by-two matrix kernel
\begin{equation}
[\boldsymbol{K}] (u,v) =  \begin{pmatrix}
 K_{11}(u,v)  &    K_{12}(u,v)  \\   K_{21}(u,v)  & K_{22}(u,v) 
\end{pmatrix}  ,
\end{equation} 
then we define its Fredholm Pfaffian on the domain $X$ as 
\begin{align}\label{eqn:FP1}
\text{Pf}& ( \boldsymbol{J}+  P_X \boldsymbol{K}  P_X) \\&
= \text{Pf}\left( \begin{pmatrix}
0  &     1 \\ -  1   & 0
\end{pmatrix}  +  \begin{pmatrix}
 P_X K_{11}   P_X &     P_X K_{12}   P_X  \\   P_X K_{21}   P_X   &  P_X K_{22}   P_X 
\end{pmatrix}  \right) 
\\&= \sum_{n=0}^\infty \frac{1}{n!} \left( \prod_{j=1}^n \int_X \rmd z_j   \right) \text{Pf} \begin{pmatrix}
 K_{11}(z_i,z_j)   &    K_{12}(z_i,z_j)  \\   K_{21}(z_i,z_j)     & K_{22}(z_i,z_j)  
\end{pmatrix}_{i,j=1}^n ,
\end{align}
where we introduced the two-by-two kernel
\begin{equation}
\boldsymbol{J} =  \begin{pmatrix}
0  &     1 \\ -  1   & 0
\end{pmatrix}  \epp
\end{equation}
The connection between Fredholm Pfaffian and Fredholm determinant is given by
\begin{equation}
\text{Det}( \boldsymbol{I} -  \boldsymbol{J}   \boldsymbol{K} ) = \text{Pf}( \boldsymbol{J} + \boldsymbol{K} )^2 .
\end{equation}

\subsection{Derivation of the formula \eqref{mainresult1}, \eqref{mainresult2} in the text}
\label{app:derpfaff} 

The expression \eqref{EqT:gbs02} for the generating function can be rewritten as
a Fredholm pfaffian:
\bea \label{FPdef}
\!\!\! \!\!\!  && g_b(s) =  {\rm Pf}[ {\bf J} + {\bf K} ]  \qquad {\bf J}=
\left(\begin{array}{cc}
0 & I  \\
- I & 0
\end{array} \right), \qquad  {\bf K}=
\left(\begin{array}{cc}
K_{11} & 0  \\
0 & - K_{22}
\end{array} \right)
\eea
with $K_{11}(v_1,v_2)=f(v_1,v_2):=f_1(v_1,v_2)+f_2(v_1,v_2)$ and $K_{22}(v_1,v_2) = 2 \delta'(v_1-v_2)$ both antisymmetric kernels. Here we assume implicitly that all integrations are
on $v_j \in [0,+\infty[$, i.e. we do not write the projectors explicitly.
The above form has similarities to the one obtained for the flat KPZ problem 
in Section 6.5 of Ref. \cite{we-flatlong}. 
The main difference is that 
here there is no off-diagonal part $K_{12}=0$, which makes the problem simpler (as discussed there, there
are two definitions of the ordering in a FP, which accounts for the minus sign.). We now follow the
same steps as in Section 6.5 of Ref.~\cite{we-flatlong}, i.e. the square of a FP is
a Fredholm determinant:
\bea
\!\!\! \!\!\! \!\!\! \!\!\! \!\!\! \!\!\! \!\!\! \!\!\! \!\!\!  && g_b(s)^2 =  {\rm Pf}[ {\bf J} + {\bf K} ]^2 = {\rm Det}[ {\bf I} - {\bf J}  {\bf K}] =  {\rm Det } \left(\begin{array}{cc}
I & K_{22}  \\
K_{11} & I
\end{array} \right) = {\rm Det}[ I - K_{22} K_{11} ]
\eea 
where we have defined:
\bea
&& (K_{22} K_{11})(v_1,v_2) = - {\cal K}(v_1,v_2) + 2 \delta(v_1) K_{11}(0,v_2) \\
&& {\cal K}(v_1,v_2) = -2 \partial_{v_1} K_{11}(v_1,v_2) = -2 \partial_{v_1} f(v_1,v_2) 
\eea 
It is shown in Section 6.5 of Ref. \cite{we-flatlong} that:
\bea
{\rm Det}[ I - K_{22} K_{11} ]  =  {\rm Det}[ I + {\cal K}  ]
\eea 
Hence we arrive at the main result \eqref{mainresult1}, \eqref{mainresult2} in the text. 

%
%
%

\section{Analytical continuations of the norm for arbitrary complex $m$ and $\ell$} \label{app:continuations}

We recall that $B_{k,m}=b_{k,m} H_{k,m}$ where, for integer $m$, using $(x)_p=(-1)^p (-x-p+1)_p$,
$b_{k,m}$ is a polynomial which can be rewritten in several equivalent ways
\bea \label{bformula} 
&& b_{k,m}= 2 m^2 4^m  S_{k,m} a_{k,m} 
=  4^m i k (i k + 1 - \frac{m}{2})_{m-1} (- i k + \frac{1-m}{2})_m \\
&& = 4^m (-1)^m i k (i k + 1 - \frac{m}{2})_{m-1} (i k + \frac{1-m}{2})_m \\
&& =  4 k^2 (1-2 i k)_{m-1} (1+2 i k)_{m-1} = (-1)^{m} 2 i k (2 i k-m+1)_{m-1} (2 i k)_{m} \nn
\eea 
each of them leads to an analytical continuation by replacing $(x)_n=\Gamma(x+n)/\Gamma(x)$.
For instance the last two give
\bea
&& b^1_{k,m}=  (-1)^m 2 i k \frac{\Gamma(2 i k + m)}{\Gamma(2 i k - m + 1)} \\
&& b^2_{k,m}=  \frac{2 k}{\pi} \sinh(2 \pi k)  \Gamma(m+ 2 i k) \Gamma(m- 2 i k) 
\eea
Note that the second has a $(-1)^m$ term compared to the first. These factors are important
since they allow to use the MB formula. 
Similarly,
\be
H_{k,m} =  \frac{(-1)^m b^{2 m}}{
\left(- b+\frac{1-m}{2}+ i k\right)_m \left(b+\frac{1-m}{2} + i k\right)_m} \\
\ee
has several analytical continuations. The first one is 
\be
 H^1_{k,m} = (-1)^m b^{2m} \frac{\Gamma\left( { b  + i k- \frac{m-1}{2}}\right)  \Gamma\left( { -b  + i k- \frac{m-1}{2}}\right) }{ \Gamma\left( { b  + i k +\frac{m+1}{2}}\right)  \Gamma \left( { -b  + i k +\frac{m+1}{2}}\right)} 
\ee
But, for instance there are two others
\bea \label{Eq:AnalyticalContinuationH2}
\fl H^2_{k,m} =  \frac{b^{2m}}{ (b + \frac{1-m}{2} - i k)_m (b + \frac{1-m}{2} + i k)_m }= 
b^{2m} \frac{\Gamma\left(b + \frac{1-m}{2} - i k \right)  \Gamma\left(b + \frac{1-m}{2} + i k\right)}{ 
\Gamma\left(b + \frac{1+m}{2} - i k \right)  \Gamma\left(b + \frac{1+m}{2} + i k\right)}
\eea
and
\bea \label{Eq:AnalyticalContinuationH3}
\fl H^3_{k,m} =   \frac{b^{2m}}{ (- b+\frac{1-m}{2} + i k )_m (- b+\frac{1-m}{2} - i k )_m}   = 
b^{2 m}  \frac{\Gamma(-b + \frac{1-m}{2} - i k)  \Gamma(-b + \frac{1-m}{2} + i k)}{ 
\Gamma(-b + \frac{1+m}{2} - i k)  \Gamma(-b + \frac{1+m}{2} + i k)}  \nn \\
\fl
\eea

Note that for $b \neq -1/2$ and $b \neq 1/2$ 
\bea
H^1_{k/(2 \lambda),m/\lambda} \simeq H^2_{k/(2 \lambda),m/\lambda} \simeq 
H^3_{k/(2 \lambda),m/\lambda} \simeq 1
\eea

We will therefore use the following analytical continuation for the function $B_{k,m}$
\begin{equation}
B_{k,m} = \frac{(2 i k) {b^{2m}}{}\frac{\Gamma(2i k + m)}{\Gamma(2 ik - m +1)}  \frac{\Gamma\left( { b  + i k- \frac{m-1}{2}}\right)  \Gamma\left( { -b  + i k- \frac{m-1}{2}}\right) }{ \Gamma\left( { b  + i k +\frac{m+1}{2}}\right)  \Gamma \left( { -b  + i k +\frac{m+1}{2}}\right)}  + \left( k \leftrightarrow - k  \right)}{2} \, .
\end{equation}
It can be checked that the first part of this formula reproduces the definition of $B_{k,m}$ \eqref{EqT:defBkm} for any $m \in \JN$. This part is also even in $k$ for any $m \in \JN$ but not for $m \in \JC$. In order to keep the symmetry $k \to -k$ of the integrand for any $m \in \JC$ we have thus symmetrized the analytical continuation.

\end{appendices}
\newpage

\input{SortedBibliography.tex}

\end{document}

%% file: SortedBibliography.tex
\section*{References}
\bibliographystyle{iopart-num}